\documentclass[12pt]{article}
\usepackage{graphicx}

\newcommand{\imp}{\mbox{\boldmath $p$}}
\newcommand{\kbf}{\mbox{\boldmath $k$}}

\newcommand{\cent}{\centerline}
\newcommand{\vs}{\vspace*}

\newcommand{\SiO}{{\rm SiO}}
\newcommand{\crm}{{\rm c}}
\newcommand{\orm}{{\rm o}}
\newcommand{\e}{{\rm e}}
\newcommand{\Xrm}{{\rm X}}
\newcommand{\Mrm}{{\rm M}}
\newcommand{\mrm}{{\rm m}}

\newcommand{\ug}{\; = \;}

\newcommand{\equ}{\; \equiv \;}

\newcommand{\text}{\rm}

\newcommand{\drm}{{\rm d}}

\newcommand{\infi}{\infty}

\newcommand{\la}{\lambda}

\newcommand{\ra}{\rightarrow}

\newcommand{\al}{\alpha}
\newcommand{\be}{\beta}
\newcommand{\ze}{\zeta}

\newcommand{\bb}{\begin{equation}}
\newcommand{\ee}{\end{equation}}
\newcommand{\bega}{\begin{eqnarray}}
\newcommand{\ega}{\end{eqnarray}}
\newcommand{\begae}{\begin{eqnarray*}}
\newcommand{\egae}{\end{eqnarray*}}

\newcommand{\h}{\hspace*{4ex}}
\newcommand{\dis}{\displaystyle}

\newcommand{\Om}{\Omega}
\newcommand{\om}{\omega}

\begin{document}

\

\hfill{NSF-ITF-02-93}

\begin{center}
{\large {\bf On the Localized Superluminal Solutions\\[0pt]
to the Maxwell equations$^{\: (\dag)}$}}
\footnotetext{$^{\: (\dag)}$ This reasearch was supported in part by the N.S.F.
under Grant No.PHY99-07949; by INFN and Miur (Italy); and by Fapesp
(Brasil). \ E-mail for contacts:  recami@mi.infn.it }
\end{center}

\vspace*{5mm}

\centerline{ Erasmo Recami$^{\rm a,c,d}$, Michel Zamboni-Rached$^{\rm b}$ }

\vspace*{0.2 cm}

\centerline{{\em $^{\rm a}$Facolt\`{a} di Ingegneria, Universit\`{a} statale di Bergamo,
Dalmine (BG), Italy;}}
\cent{{\em $^{\rm b}$DMO--FEEC, \ UNICAMP, \ Campinas, SP, Brasil;}}
\centerline{{\em $^{\rm c}$INFN---Sezione di Milano, Milan,
Italy; \ {\rm and}}} \centerline{{\em $^{\rm d}$Kavli Institute for Theoretical Physics,
UCSB, CA 93106, USA.}}

\vs{0.2 cm}

\centerline{\rm and}

\vs{0.4 cm}

\cent{K. Z. N\'obrega,
C\'esar A. Dartora and Hugo E. Hern\'andez-Figueroa}

\vs{0.4 cm}

\cent{{\em DMO--FEEC, Universidade Estadual de Campinas, Campinas, SP, Brasil.}}

\

\vspace*{1. cm}

{\em Abstract -} In the first part of this article (after a
sketchy theoretical introduction) the various experimental sectors
of physics in which Superluminal motions seem to appear are
briefly mentioned. In particular, a bird's-eye view is presented
of the experiments with evanescent waves (and/or tunneling
photons), and with the ``localized Superluminal solutions" (SLS)
to the wave equation, like the so-called X-shaped beams. \ In {\em the
second part} of this paper we present a series of new SLSs to the
Maxwell equations, suitable for arbitrary frequencies and
arbitrary bandwidths: some of them being endowed with finite total
energy.  Among the others, we set forth an infinite family of
generalizations of the classic X-shaped wave; and show how to
deal with the case of a {\em dispersive} medium. \ Results of this
kind may find application in other fields in which an essential
role is played by a wave-equation (like acoustics, seismology,
geophysics, gravitation, elementary particle physics, etc.). \
This e-print, in large part a review, was prepared for the special
issue on ``Nontraditional Forms of Light" of the IEEE JSTQE
(2003); and a preliminary version of it appeared as Report
NSF-ITP-02-93 (ITP, UCSB; 2002). \ Further material can be found
in the recent e-prints arVive:0708.1655v2][physics.gen-ph] and
arVive:0708.1209v1][physics.gen-ph]. \ The case of the very interesting
(and more orthodox, in a sense) {\em subluminal} Localized Waves,
solutions to the wave equations, will be dealt with in a coming paper. \
[Keywords:  Wave equation;
Wave propagation; Localized solutions to Maxwell equations;
Superluminal waves; Bessel beams; Limited-dispersion beams;
Electromagnetic wavelets; X-shaped waves; Finite-energy beams;
Optics; Electromagnetism; Microwaves; Special relativity; PACS
nos.: 03.50.De; 41.20.Jb; 83.50.Vr; 62.30.+d; 43.60.+d; 91.30.Fn;
04.30.Nk; 42.25.Bs; 46.40.Cd; 52.35.Lv]

\

\

{\large\bf 1. - Introduction.}\hfill\break

\h The question of Superluminal ($V^{2}>c^{2}$) objects or
waves\footnote{It is an old use of ours to write Superluminal with a
capital S.}, has a long story, starting perhaps in 50 B.C. with Lucretius'
{\em De Rerum
Natura} (cf., e.g., book 4, line 201: [$<<$Quone vides {\em citius} debere
et longius ire/ Multiplexque loci spatium transcurrere eodem/ Tempore {\em
quo Solis} pervolgant {\em lumina} coelum?$>>$]). \ Still in
pre-relativistic times, one meets various related works, from those by
J.J.Thomson to the papers by A.Sommerfeld. \ With Special
Relativity, however, since 1905 the conviction spread over that the speed $c$
of light in vacuum was the upper limit of any possible speed. For
instance, R.C.Tolman in 1917 believed to have shown by his ``paradox'' that
the existence of particles endowed with speeds larger than $c$ would have
allowed sending information into the past. Such a conviction blocked
any research about Superluminal speeds. Our problem started to be tackled
again essentially in the fifties
and sixties, in particular after the papers[1] by E.C.George Sudarshan
et al., and, later on,[2] by E.Recami, R.Mignani et al., as well as by
H.C.Corben and others (to confine
ourselves to the {\em theoretical} researches). \ The first experiments
looking for faster-than-light objects were performed by T.Alv\"{a}ger
et al.[1].

\h Superluminal objects (particles or wavepackets) were called tachyons, T, by G.Feinberg,
from the Greek word $\tau \alpha \chi {\acute{\upsilon}} \varsigma$, quick
[and this induced us in 1970 to coin the term bradyon, B, for ordinary
subluminal ($v^2<c^2$) objects, from the Greek word $\beta \rho \alpha
\delta {\acute{\upsilon}} \varsigma$, slow].  Finally, objects travelling
exactly at the speed of light are called ``luxons".

\h  In recent years, terms as ``tachyon'' and ``superluminal''
fell unhappily into the (cunning, rather than crazy) hands of
pranotherapists and mere cheats, who started squeezing money out
of simple-minded people; for instance by selling plasters (!) that
should cure various illnesses by ``emitting tachyons''... \ Here
we are dealing, however, with Superluminal waves or objects since
at least four different experimental sectors of physics seem to
indicate the actual existence, in a sense, of Superluminal motions
[thus confirming some long-standing theoretical predictions[3]].

\h In {\em the first part} of this article (after a sketchy,
non-technical theoretical introduction, which might be useful
since it enlightens an original, scarcely known
approach\footnote{For a more detailed reviews, see P.W.Milonni, in
ref.[22] below, and the recent e-prints
arVive:0708.1655v2][physics.gen-ph] and
arVive:0708.1209v1][physics.gen-ph].}) we briefly mention the
various experimental sectors of physics in which Superluminal
motions seem to appear. In particular, a bird's-eye view is
presented of the experiments with evanescent waves (and/or
tunneling photons), and  with the ``localized Superluminal
solutions" (SLS) to the wave equation, like the so-called X-shaped
beams; the shortness of this review is compensated by a number of
references, sufficient in some cases to provide the interested
readers with reasonable bibliographical information. \ In the
second part of this paper ---after having constructed an infinite
family of generalizations of the classic X-shaped wave, and
having briefly discussed the behavior of some {\em finite}
total-energy SLSs--- we propose a series of new SLSs to the
Maxwell equations, suitable for arbitrary frequencies and
bandwidths, and show moreover how to deal with
the case of a {\em dispersive} medium.\\

\h Let us state first that special relativity (SR), abundantly
verified by experience, can be built on two simple, natural Postulates: \ 1)
that the laws (of electromagnetism and mechanics) be valid not only for a
particular observer, but for the whole class of the ``inertial" observers: \
2) that space and time be homogeneous and space be moreover isotropic. \
From these Postulates one can theoretically {\em infer} that one, and only
one, {\em invariant} speed exists: and experience tells us such a speed to
be the one, $c$, of light in vacuum; indeed, ordinary light possesses the
peculiar feature of presenting always the same speed in vacuum, even when we
run towards or away from it. \ It is just that feature, of being invariant,
that makes quite exceptional the speed $c$: no slower, nor faster speed
can enjoy the same property.

\begin{figure}[!h]
\begin{center}
 \scalebox{.75}{\includegraphics{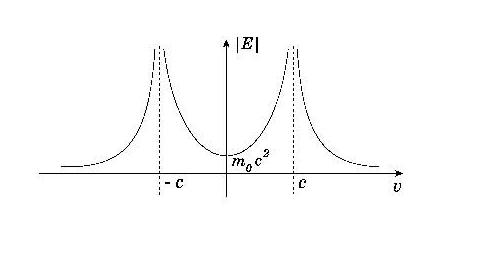}}
\end{center}
\caption{Energy of a free object as a function of its speed.[2-4]}
\label{fig1}
\end{figure}

\h  Another (known) consequence of our Postulates is that the
total energy of an ordinary particle increases when its speed $v$ increases,
tending to infinity when $v$ tends to $c$. Therefore, infinite forces would
be needed for a bradyon (= slower than light object) to reach the speed $c$.
However, as speed $c$ photons exist which are born live and die always at
the speed of light (without any need of accelerating from rest to the light
speed), so objects can exist[4] always endowed with speeds
$V$ larger than $c$ (see Fig.\ref{fig1}). \ [This circumstance has been picturesquely
illustrated by George Sudarshan (1972) with reference to an imaginary
demographer studying the population patterns of the Indian subcontinent:
$<<$Suppose a demographer calmly asserts that there are no people North of the
Himalayas, since none could climb over the mountain ranges! That would be an
absurd conclusion. People of central Asia are born there and live there:
they did not have to be born in India and cross the mountain range. So with
faster-than-light particles$>>$.] \ Let us add that, still starting from the
above two Postulates (besides a third postulate, even more obvious), the
theory of relativity can be generalized[3,4] in such a way to
accommodate also Superluminal objects; a large part of such an extension
being contained in a series of works which date back to the
sixties--seventies. \ Also within ``Extended Relativity''[3] the
speed $c$, besides being invariant, is a limiting velocity: but every
limiting value has two sides, and one can a priori approach it both from
the left and from the right.

\begin{figure}[!h]
\begin{center}
 \scalebox{0.75}{\includegraphics{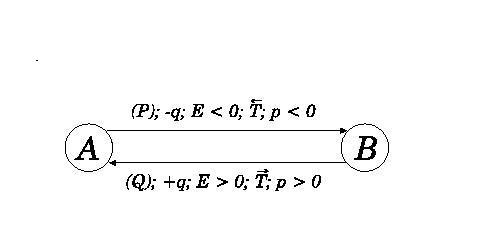}}
\end{center}
\caption{Depicting the ``switching rule" (or reinterpretation principle) by
Stueckelberg-Feynman-Sudarshan-Recami[3-5]: \ $Q$ will appear as the
antiparticle of P. \ See the text.}
\label{fig2}
\end{figure}

\h  Actually, the ordinary formulation of SR is restricted too
much. For instance, {\em even leaving Superluminal speeds aside},
it can be easily so widened as to include antimatter[5]. Then, one
finds space-time to be a priori populated by normal particles P,
{\em and} by dual particles Q ``which travel backwards in time
carrying negative energy'': The latter shall appear to us as
antiparticles, i.e., as particles ---regularly traveling forward
in time with positive energy, but--- with all their ``additive''
charges (e.g., the electric charge) reversed in sign: see
Fig.\ref{fig2}. and refs.[3-5]. \ To clarify this point, one could
here recall only what follows: We, macroscopic observers, have to
move in time along a single, well-defined direction, to such an
extent that we cannot even see a motion backwards in time...; and
every object like Q, travelling backwards in time (with negative
energy), will be {\em necessarily} reinterpreted by us as an
anti-object, with opposite charges but travelling forward in time
(with positive energy): cf. again Fig.2 and refs.[3-5].

\h But let us forget about antimatter and go back to ``tachyons"
(= faster-than-light waves or objects). A strong
objection against their existence is based on the opinion that by
tachyons it would be possible to send signals into the past, owing to the fact
that a tachyon T which, say, appears to a first observer $O$ as emitted
by A and absorbed by B, can appear to a second observer $O^{\prime}$ as a
tachyon T' which travels backwards in time with negative energy. However, by
applying the same ``reinterpretation rule" or switching procedure seen
above (Fig.2), T' will appear to the new observer
$O^{\prime}$ just as an antitachyon ${\overline{{\rm T}}}$ emitted by B and
absorbed by A, and therefore traveling forward in time, even if in the
contrary {\em space} direction.\footnote{In simple words, the reinterpretation
rule implies that signals are carried only by objects which appear to be
endowed with positive energy.} In such a way, every travel towards the
past, and every negative energy, disappear: Cf. refs.[3-5].

Starting from this observation, it is possible to solve[5] the so-called
causal paradoxes associated with Superluminal motions: paradoxes which
result to be the more instructive, the more sophisticated they
are; \ but that cannot be re-examined here (some of them having been
proposed by R.C.Tolman, J.Bell, F.A.E.Pirani, J.D.Edmonds and others).[6,3]
\ [Let us only mention here the following. \ The reinterpretation principle,
according to which, in simple words, signals are carried only by objects
which appear to be endowed with positive energy does eliminate any
information transfer backwards in time; but this has a price: That of
abandoning the ingrained conviction that the judgement about what is cause
and what is effect be independent of the observer. In fact, in the case
examined above, the first observer $O$ considers the event at A to be the
cause of the event at B. \ By contrast, the second observer $O^{\prime}$
will consider the event at B as causing the event at A. \ All the observers
will however see the cause to happen {\em before} its effect.]\footnote{Taking
new objects or entities into consideration always
forces us to a criticism of our prejudices. If we require the phenomena to
obey the {\em law} of (retarded) causality with respect to all the
observers, then we cannot demand also the {\em description} ``details'' of
the phenomena to be invariant: namely, we cannot demand in that case also the
invariance of the ``cause'' and ``effect'' labels.[6,2]}

\h  Let us observe that last century theoretical physics led us in a natural
way to suppose the existence of various types of objects: magnetic monopoles,
quarks, strings, tachyons, besides black-holes: and various sectors of
physics could not go on without them, even if the existence of none of them
is certain (also because attention has not yet been paid to some links
existing among them: e.g., a Superluminal electric charge is expected to
behave as a magnetic monopole; and a black-hole a priori can be the source
of tachyonic matter).\ According to Democritus of Abdera, everything that
was thinkable without meeting contradictions had to exist somewhere in the
unlimited universe. This point of view ---which was given by M.Gell-Mann the
name of ``totalitarian principle''--- was later on expressed (T.H.White) in
the humorous form ``Anything not forbidden is compulsory''...

\h Let us add here the information that the case of the very interesting
(and more orthodox, in a sense) {\em subluminal} Localized Waves,
solutions to the wave equations, will be dealt with in a coming paper.

\

{\large\bf 2. A glance at the experimental status-of-the-art}.\hfill\break

\h Extended Relativity can allow a better understanding of many
aspects also of {\em ordinary} physics, even if tachyons would
not exist in our cosmos as asymptotically free objects. \ As already said,
however, at least three or four different experimental sectors of physics
seem to suggest the possible existence of faster-than-light motions; as
our first aim, in the following we set forth some information (mainly
bibliographical) about the experimental results obtained in a couple of
those different physics sectors, with a mere mention of the others.

\

{\bf A)} \ {\em Neutrinos} -- First: A long series of experiments, started
in 1971, seems to show that the square ${m_{0}}^{2}$ of the mass $m_{0}$ of
muon-neutrinos, and more recently of electron-neutrinos too, is
negative; which, if confirmed, would mean that (when using a na\"{i}ve
language, commonly adopted) such neutrinos possess an ``imaginary mass'' and
are therefore tachyonic, or mainly tachyonic.[7,3] \ [In Extended
Relativity, the dispersion relation for a free Superluminal object becomes \
$\om^2-\kbf^2=-\Om^2$, \ or \ $E^2-\imp^2=-m_\orm^{2}$, \ and there is
{\em no} need therefore of imaginary masses...].\footnote{Let us put $c=1$,
{\em whenever convenient,} throughout this paper.}

\

{\bf B)} \ {\em Galactic Micro-quasars} -- Second: As to the {\em apparent}
Superluminal expansions observed in the core of quasars[8] and, recently, in
the so-called galactic microquasars[9], we shall not deal here with that
problem, too far from the other topics of this paper: without mentioning
that for those astronomical observations there exist orthodox
interpretations, based on ref.[10], that are accepted by the majority
of astrophysicists. \ For a theoretical discussion, see ref.[11].
Here, we may be interested in the fact that simple geometrical considerations
in Minkowski space show that a {\em single} Superluminal light source would
appear[11,3]: \ (i) initially, in the ``optical boom'' phase (analogous to the
acoustic ``boom'' produced by a plane traveling with constant supersonic
speed), as an intense source which suddenly comes into view; and that \ (ii)
afterwards seem to split into TWO objects receding one from the other with
speed \ $V>2c$ [all of this being similar to what is actually
observed, according to refs.[9]].

\

\begin{figure}[!h]
\begin{center}
 \scalebox{.75}{\includegraphics{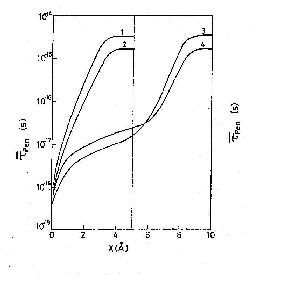}}
\end{center}
\caption{Behaviour of the average ``penetration time" (in seconds) spent by
a tunneling wavepacket, as a function of the penetration depth (in
{\aa}angstroms) down a potential barrier (from Olkhovsky et al., ref.[12]).
According to the predictions of quantum mechanics, the wavepacket speed
inside the barrier increases in an unlimited way for opaque barriers; and the
total tunneling time does {\em not} depend on the barrier width[12].}
\label{fig3}
\end{figure}

\begin{figure}[!h]
\begin{center}
 \scalebox{.65}{\includegraphics{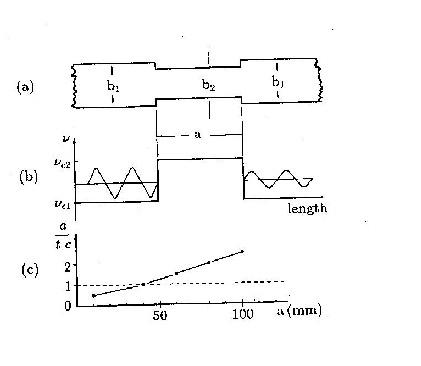}}
\end{center}
\caption{Simulation of tunneling by experiments with evanescent classical
waves (see the text), which were predicted to be Superluminal also on the
basis of Extended Relativity[3,4]. The figure shows one of the measurement
results in refs.[15]; that is, the average beam speed while crossing
the evanescent region ( = segment of undersized waveguide, or ``barrier")
as a function of its length. As theoretically predicted[19,12], such an
average speed exceeds $c$ for long enough ``barriers".}
\label{fig4}
\end{figure}

{\bf C)} \ {\em Evanescent waves and ``tunneling photons''} -- Third:
Within quantum mechanics, it had been shown that the tunneling time does
not depend on the width of the barrier, in the case of
opaque barriers (``Hartman effect'')[12]. This implies Superluminal and
arbitrarily large (group) velocities $V$ inside long enough barriers: see
Fig.\ref{fig3} \ Experiments that may verify this prediction by, say, electrons are
difficult. Luckily enough, however, the Schroedinger equation in the
presence of a potential barrier is mathematically identical to the Helmholtz
equation for an electromagnetic wave propagating, for instance, down a
metallic waveguide along the $x$-axis: as shown, e.g., by R.Chiao et
al.[13]; \ and a barrier height $U$ bigger than the electron energy $E$ corresponds
(for a given wave frequency) to a waveguide of transverse size lower than a
cut-off value. A segment of ``undersized" guide ---to go on with our example---
does therefore behave as a barrier for the wave (photonic barrier)[13],
as well as any other
photonic band-gap filters. \ The wave assumes therein ---like an electron
inside a quantum barrier--- an imaginary momentum or wave-number and gets,
as a consequence, exponentially damped along $x$. In other words, it
becomes an {\em evanescent} wave (going back to normal propagation, even
if with reduced amplitude, when the narrowing ends and the guide returns to
its initial transverse size). \ Thus, a tunneling experiment can be
simulated[13] by having
recourse to evanescent waves (for which the concept of group velocity can be
properly extended[14]). The fact that evanescent waves travel with
Superluminal speeds (cf.,e.g., Fig.\ref{fig4}) has been actually {\em verified} in a
series of famous experiments.

\h Namely, various experiments, performed since 1992 onwards by
R.Chiao, P.G.Kwiat and A.Steinberg's group at Berkeley[15], by
G.Nimtz et al. at Cologne[16], by A.Ranfagni and colleagues at
Florence[17], and by others at Vienna, Orsay, Rennes[17], verified
that ``tunneling photons" travel with Superluminal group
velocities. \ Let us add that also Extended Relativity had
predicted[19] evanescent waves to be endowed with faster-than-$c$
speeds; the whole matter appears to be therefore theoretically
self-consistent. \ The debate in the current literature does not
refer to the experimental results (which can be correctly
reproduced by numerical elaborations[20,21] based on Maxwell
equations only), but rather to the question whether they allow, or
do not allow, sending signals or information with Superluminal
speed[22,21,14].

\h In the abovementioned experiments one meets a substantial attenuation
of the considered pulses during tunneling (or during propagation in an
absorbing medium). However, by employing a ``gain doublet", it has been recently
reported the observation of undistorted pulses propagating with Superluminal
group-velocity with a {\em small} change in amplitude.[23]

\begin{figure}[!h]
\begin{center}
 \scalebox{.75}{\includegraphics{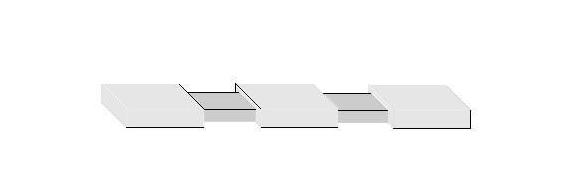}}
\end{center}
\caption{The very interesting experiment along a metallic waveguide with
TWO barriers (undersized guide segments), i.e., with two evanescence
regions[24]. See the text.}
\label{fig5}
\end{figure}

\h  Let us emphasize that some of the most interesting experiments
of this series seem to be the ones with two ``barriers" (e.g.,
with two gratings in an optical fiber, or with two segments of
undersized waveguide separated by a piece of normal-sized
waveguide: Fig.\ref{fig5}). For suitable frequency bands ---i.e.,
for ``tunneling" far from resonances---, it was found that the
total crossing time does not depend on the length of the
intermediate (normal) guide: namely, that the beam speed along it
is infinite[24,25]. \ This agrees with what predicted by Quantum
Mechanics for the non-resonant tunneling through two successive
opaque barriers (the tunneling phase time, which depends on the
entering energy, has been shown to be {\em independent} of the
distance between the two barriers[26]); something that has been
accepted and generalized in Aharonov et al.[26] \ Such a
prediction has been verified a second time, taking advantage of
the circumstance that quite interesting evanescence regions can be
constructed in the most varied manners, like by means of different
photonic band-gap materials or gratings (it being possible to use
from multi-layer dielectric mirrors, or semiconductors, to
photonic crystals...). And indeed a very recent confirmation came
---as already mentioned--- from an experiment having recourse to
two gratings in an optical fiber.[25]

\h  We cannot skip a further, indeed delicate, topic, since the last
experimental contribution to it[23] aroused large interest. \
Even if in Extended Relativity all the ordinary causal paradoxes seem to
be solvable[3,6], nevertheless one has to bear in mind that (whenever it
is met an object, ${\cal O}$, traveling with Superluminal speed) one may have
to deal with negative contributions to the {\em tunneling times\/}[27,12]:
and this should not be regarded as unphysical. In fact, whenever an ``object''
(particle, electromagnetic pulse,,...) ${\cal O}$ {\em overcomes} the
infinite speed[3,6] with respect to a certain observer, it will afterwards
appear to the same observer as the ``{\em anti}-object'' $\overline{{\cal O}}
$ traveling in the opposite {\em space} direction[3,6]. \ For instance,
when going on from the lab to a frame ${\cal F}$ moving in the {\em same}
direction as the particles or waves entering the barrier region, the object
${\cal O}$ penetrating through the final part of the barrier (with almost
infinite speed[12,21,26,27], like in Figs.3) will appear in the frame
${\cal F}$ as an anti-object $\overline{{\cal O}}$ crossing that portion
of the barrier {\em in the opposite space-direction\/}[3,6]. In the new
frame ${\cal F}$, therefore, such anti-object $\overline{{\cal O}}$ would
yield a {\em negative} contribution to the tunneling time: which could
even result, in total, to be negative. \ For any clarifications, see
refs.[28]. \ What we want to stress here is that the appearance of such
negative times is predicted by Relativity itself, on the basis of the
ordinary postulates[3,6,21,28]. \ (In the case of a non-polarized beam, the
wave anti-packet coincides with the initial wave packet; if a photon is
however endowed with helicity $\lambda =+1$, the anti-photon will bear the
opposite helicity $\lambda =-1$). \ From the theoretical point of view,
besides refs.[3,6,12,21,27,28], see refs.[29]. \ On the (quite interesting)
experimental side, see papers [30], the first one having already been
mentioned above.

\h  Let us {\em add} here that, via quantum interference effects
it is possible to obtain dielectrics with refraction indices very rapidly
varying as a function of frequency, also in three-level atomic systems, with
almost complete absence of light absorption (i.e., with quantum induced
transparency)[31]. \ The group velocity of a light pulse propagating in
such a medium can decrease to very low values, either positive or negatives,
with {\em no} pulse distortion. \ It is known that experiments have been
performed both in atomic samples at room temperature, and in Bose-Einstein
condensates, which showed the possibility of reducing the speed of light to
a few meters per second. \ Similar, but negative group velocities, implying
a propagation with Superluminal speeds thousands of time higher than the
previously mentioned ones, have been recently predicted also in the presence
of such an ``electromagnetically induced transparency'', for light moving in
a rubidium condensate.[32]  Finally, let us recall that faster-than-$c$
propagation of
light pulses can be (and was, in same cases) observed also by taking
advantage of anomalous dispersion near an absorbing line, or nonlinear and
linear gain lines ---as already seen---, or nondispersive dielectric media,
or inverted two-level media, as well as of some parametric processes in
nonlinear optics (cf., e.g., G.Kurizki et al.[30]).

\

{\bf D)} \ {\em Superluminal Localized Solutions (SLS) to the wave
equations. The ``X-shaped waves"} -- The fourth sector (to leave aside the
others) is not less important. It came into fashion again, when it was
rediscovered that any wave equation ---to fix the ideas, let us think
of the electromagnetic
case--- admit also solutions as much sub-luminal as Super-luminal (besides
the ordinary waves endowed with speed $c/n$). \ Actually,
starting with the pioneering work by H.Bateman, it had slowly become known
that all homogeneous wave equations (in a general sense: scalar,
electromagnetic, spinorial,...) admit wavelet-type solutions with
sub-luminal group velocities[33]. \ Subsequently, also Superluminal
solutions started to be written down, in refs.[34] and, independently, in
refs.[35] (in one case just by the mere application of a Superluminal
Lorentz ``transformation"[3,36]).

\begin{figure}[!h]
\begin{center}
 \scalebox{.75}{\includegraphics{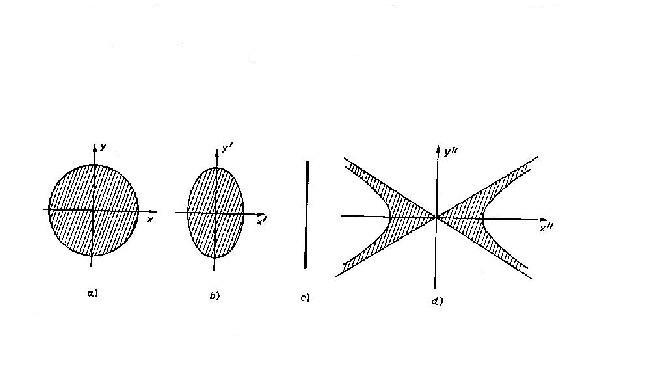}}
\end{center}
\caption{An intrinsically spherical (or pointlike, at the limit) object
appears in the vacuum as an ellipsoid contracted along the motion direction
when endowed with a speed $v<c$. \ By contrast, if endowed with a speed
$V>c$ (even if the $c$-speed barrier cannot be crossed, neither from the
left nor from the right), it would appear[37] no longer as a particle, but
rather as an ``X-shaped" wave[37] traveling rigidly (namely, as occupying
the region delimited by a double cone and a two-sheeted hyperboloid ---or
as a double cone, at the limit--, moving Superluminally and without
distortion in the vacuum, or in a homogeneous medium).}
\label{fig6}
\end{figure}

\begin{figure}[!h]
\begin{center}
 \scalebox{.65}{\includegraphics{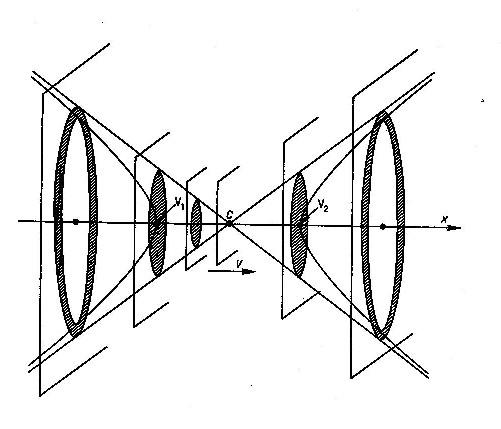}}
\end{center}
\caption{Here we show the intersections of an ``X-shaped wave"[37] with
planes orthogonal to its motion line, according to Extended Relativity``[2-4].
The examination of this figure suggests how to construct a simple dynamic
antenna for generating such localized Superluminal waves (such an antenna
was in fact adopted, independently, by Lu et al.[39] for the production
of such non-dispersive beams).}
\label{fig7}
\end{figure}

\h  An important feature of some new solutions of these (which
attracted much attention for possible applications) is that they
propagate as localized, non-dispersive pulses: namely, as
``undistorted progressive waves'', according to the Courant and
Hilbert's[33] terminology. It is easy to realize the practical
importance, for instance, of a radio transmission carried out by
localized beams, independently of their being sub- or
Super-luminal. \ But non-dispersive wave packets can be of use
even in theoretical physics for a reasonable representation of
elementary particles[37], and so on. \ Within Extended Relativity
since 1980 it had been found[38] that ---whilst the simplest
subluminal object conceivable is a small sphere, or a point as its
limit--- the simplest Superluminal objects results by contrast to
be (see refs.[38], and Fig.\ref{fig6} and Fig.\ref{fig7} of this
present paper) an ``X-shaped'' wave, or a double cone as its
limit, which moreover travels without deforming ---i.e.,
rigidly--- in a homogeneous medium[3]. \ It is not without meaning
that the most interesting localized solutions happened to be just
the Superluminal ones, and with a shape of that kind. \ Even more,
since from Maxwell equations under simple hypotheses one goes on
to the usual {\em scalar} wave equation for each electric or
magnetic field component, one can expect the same solutions to
exist also in the field of acoustic waves, and of seismic waves
(and of gravitational waves too). \ Actually, such beams (as
suitable superpositions of Bessel beams[39]) were mathematically
constructed for the first time, by Lu et al.[40], {\em in
acoustics\/}: and were then called ``X-waves'' or rather X-shaped
waves.

\begin{figure}[!h]
\begin{center}
 \scalebox{.65}{\includegraphics{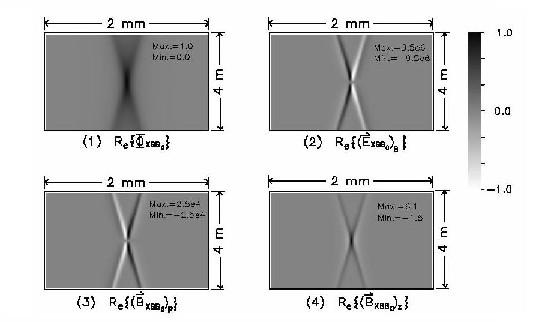}}
\end{center}
\caption{Theoretical prediction of the Superluminal localized ``X-shaped"
waves for the electromagnetic case (from Lu, Greenleaf and Recami[40], and
Recami[40]).}
\label{fig8}
\end{figure}

\begin{figure}[!h]
\begin{center}
 \scalebox{.75}{\includegraphics{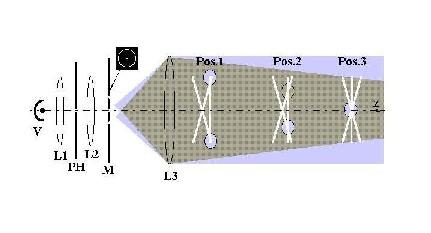}}
\end{center}
\caption{Scheme of the experiment by Saari et al., who announced (PRL of 24
Nov.1997) the production in optics of the beams depicted in Fig.8: In this
figure one can see what shown by the experiment, i.e., that the Superluminal
``X-shaped" waves which run after and catch up with the plane waves (the
latter regularly traveling with speed $c$). An analogous experiment has
been performed with microwaves at Florence by Mugnai, Ranfagni and Ruggeri
(PRL of 22 May 2000).}
\label{fig9}
\end{figure}

\h  It is more important for us that the X-shaped waves have been
indeed produced in experiments both with acoustic and with
electromagnetic waves; that is, X-beams were produced which, in
their medium, travel undistorted with a speed larger than sound,
in the first case, and than light, in the second case. \ In
acoustics, the first experiment was performed by Lu et al.
themselves[41] in 1992. \ In the electromagnetic case, certainly
more intriguing, Superluminal localized X-shaped solutions were
first mathematically constructed by us (cf., e.g., our
Fig.\ref{fig8} taken from refs.[42]), and later on experimentally
produced by Saari et al.[43] in 1997 by visible light
(Fig.\ref{fig9}), and more recently by Mugnai, Ranfagni and
Ruggeri at Florence by microwaves[44], and, again in optics but
exploting this time the response of a nonlinear medium, by Di
Trapani et al.[44].\ Further experimental activity is in progress;
while in the theoretical sector the activity has been not less
intense, in order to build up ---for example--- new analogous
solutions with finite total energy or more suitable for high
frequencies (and with adjustable bandwidth), on the one hand[45],
and localized solutions Superluminally propagating even along
normal waveguides, on the other hand[46]: and we are going to
contribute, below, just to questions of this type.

\h  Before going on, let us eventually mention the problem of
producing an X-shaped Superluminal wave like the one in Fig.7, but
truncated ---of course-- in space and in time (by the use of a
finite antenna, radiating for a finite time: a {\em dynamic}
antenna, in general): in such a situation, the wave will keep its
localization and Superluminality only along a certain ``depth of
field'', decaying abruptly[40,42] afterwards.\footnote{We can
become convinced about the possibility of realizing it, by imaging
the simple ideal case of a negligibly sized Superluminal source
$S$ endowed with speed $V>c$ in vacuum and emitting
electromagnetic waves $W$ (each one traveling with the invariant
speed $c$). The electromagnetic waves will result to be internally
tangent to an enveloping cone $C$ having $S$ as its vertex, and as
its axis the propagation line $x$ of the source[3]. \ This is
analogous to what happens for a plane that moves in the air with
constant supersonic speed. \ The waves $W$ interfere negatively
inside the cone $C$, and constructively only on its surface. \ We
can place a plane detector orthogonally to $x$, and record
magnitude and direction of the $W$ waves that hit on it, as
(cylindrically symmetric) functions of position and of time. \ It
will be enough, then, to replace the plane detector with a plane
antenna which {\em emits} ---instead of recording--- exactly the
same (axially symmetric) space-time pattern of waves $W$, for
constructing a cone-shaped electromagnetic wave $C$ that will
propagate with the Superluminal speed $V$ (of course, without a
source any longer at its vertex): \ even if each wave $W$ travels
with the invariant speed $c$. \ For further details, see the first
of refs.[42].} \ Actually, such localized Superluminal beams
appear to keep their good properties only as long as they are fed
by the waves arriving (with speed $c$) from the antenna: Taking
account of the time needed for fostering such Superluminal pulses
(i.e., for the arrival of the feeding speed-$c$ waves coming from
the antenna). one concludes that these localized Superluminal
beams are probably unable to transmit {\em information} faster
than $c$. However, they don't seem to have anything to do with the
illusory ``scissors effect'', since they carry energy-momentum
Superluminally along their field depth [for instance, they can get
two (tiny) detectors at a distance $L$ to click after a time {\em
smaller} than $L/c$]. \ As we mentioned above, the existence of
all these X-shaped Superluminal (or ``Super-sonic'') waves,
together with the Superluminality of evanescent waves, are
compatible with (Extended) Relativity: a theory based, let us
repeat, on the ordinary postulates of SR and that consequently
cannot be in contrast with any of its fundamental principles.  And
in fact all the previous results may be obtained from the Maxwell
equations (or the wave equation) only. \ It is worthwhile
mentioning, moreover, that one of the first applications of the
X-waves (which takes advantage of their propagation without
deformation) is in progress in the
field of medicine, and precisely of ultrasound scanners[47].

\

{\large\bf 3. SLSs to the Maxwell equations with arbitrary frequencies or
finite total-energy: An outline}.\hfill\break

\h As we recalled above, since many years it has been known
that localized
(non-dispersive) solutions exist to the (homogeneous, and even non-homogeneous)
wave equation[33-36], endowed with subluminal or Superluminal[42] velocities.
Particular attention has been paid to the Superluminal localized solutions,
which seem to propagate not only in vacuum but also in media with
boundaries[48], like normal-sized metallic waveguides[14] and, possibly,
optical fibers.  Such SLS have been experimentally produced.

\h However, all the analytical SLSs known to us, with one
exception[49], have a broad-band frequency spectrum starting from
zero, they being appropriate for low frequency regions only. This
fact can be a problem, because it is difficult or even impossible
to define a carrier frequency for those solutions (as well as to
use them in high frequency applications).  It is therefore
interesting to obtain exact SLSs to the wave equation with spectra
localized at higher (arbitrary) frequencies and with an adjustable
bandwidth (in other words, with a well defined carrier frequency).
In previous work of ours[49] we showed how to shift the spectrum
to higher frequencies, without dealing however with its bandwidth.
In the second part of this paper we are going to propose
analytical and exact SLSs in vacuum, whose spectra can be
localized inside any range of frequency with adjustable
bandwidths, and therefore with the possibility of choosing a well
defined carrier frequency. In this way, we can construct (without
any approximation) radio, microwave, optical, etc., localized
Superluminal waves.

\h We shall also show how, taking advantage of our methodology, we can
obtain the first analytical approximations to SLSs in {\em dispersive} media
(i.e., in media with a frequency dependent refractive index). One of the
interesting points of our approach is that all results are obtained from
simple mathematical operations on the classic ``X-wave" function.

\h Before going on, let us set forth an infinite family of
generalizations of the classic X-shaped wave; and briefly mention the
problem of the construction of {\em finite} total-energy SLSs (which is being
considered also in refs.[45]). However, We shall have to skip here
other questions, like that of obtaining SLSs which propagate
along waveguides (a question that is being considered in refs.[46]).\\

\

{\bf 3.1} -- {\bf Standard low-frequency SLSs, and generalizations}\\

\h Let us start from the axially symmetric solution (Bessel beam) to the wave
equation in vacuum, in cylindrical co-ordinates:

\

\hfill{$
\psi (\rho,z,t) \ug J_0(k_\rho \rho) \; \dis{\e^{+ik_z z} \; \e^{-i\om t}}
$\hfill} (1)

\

with the conditions

\

\hfill{$
k_\rho^2 = \dis{\frac{n^2 \om^2}{c^2} - k_z^2} \; ; \ \ \ \ \ \ k_\rho^2 \geq 0 \ ,
$\hfill} (2)

\

where $J_0$ is the zeroth-order ordinary Bessel function; quantities
$k_z \equiv k_\Vert$ and $k_{\rho} \equiv k_\perp$ (often called simply $k$, with ambiguous
notations) are the axial and the transverse wavenumber magnitude,
respectively; while $\om$ is the angular frequency; $c$ the light speed in
vacuum; and, in the vacuum case, $n=1$.

\h It is essential to stress right now that the dispersion relation (2), with
positive (but not constant, a priori) $k_\rho^2$ and real $k_z$, while
enforcing the consideration
of the truly propagating waves only (with exclusion of the evanescent ones),
does allow for both subluminal and Superluminal solutions; the latter being
the ones of interest here for us. [Conditions (2) correspond in the
($\om,k_z$) plane to confining ourselves to the region delimited by the
straight lines $\om = \pm c k_z$].

\h In this Sec.3 we shall consider, for simplicity, the vacuum case. \ A
general, axially symmetric superposition of Bessel beams (with $\Phi'$
as spectral weight-function) will therefore be:

\

\hfill{$
\Psi(\rho,z,t) \, = \, \dis{\int_{0}^{\infi}\drm k_\rho\;\int_{0}^{\infi}\drm \om \;
\int_{-\om /c}^{+\om /c}\drm k_z} \, \psi(\rho,z,t) \, \delta \left(k_\rho-
\sqrt{{{\om^2} \over {c^2}} - k_z^2} \right) \, \Phi'(\om,k_z;k_\rho) \; ,
$\hfill} (3)

\

where we put $n=1$ for simplicity's sake.

\h Let us recall also that each Bessel beam is associated with an
(``axicone") angle $\theta$, linked to its speed $V$ by the known
relations[38]:

\

\hfill{$ \tan\theta = \sqrt{V^2 -1}; \ \ \ \ \sin\theta =
\dis{{{\sqrt{V^2 -1} \over V}}}; \ \ \ \ \cos\theta = \dis{{c
\over V}} \ , $\hfill} (4)

\

where $V>c$, and more precisely $V \ra c$ when $\theta \ra 0$
while $V \ra \infi$ when $\theta \ra \pi / 2$.

\h In the first one of refs.[45], we have however shown that instead of
eq.(3) one may consider the more easily integrable Bessel beam superposition
[with $V \geq 1$]

\

\hfill{$
\begin{array}{clr}\Psi(\rho,\ze,\eta) \ug \dis{\int_{0}^{\infi}
\drm k_\rho\,\int_{0}^{\infi}
\drm \al \,\int_{0}^{\infi}\drm \be \; J_0(k_\rho \rho) \; \e^{i\al\ze} \;
\e^{-i\be\eta}} \, \times\;\;\;\;
\;\;\;\\ \;\;\;\; \\ \;\;\;\;\;\;\;\;\;\;\;\;\;\;\;\;\;\;\;\;\;\;\;\;
\times \, \delta\left(k_\rho-\sqrt{(\al^2 + \be^2)(V^2-1) + 2(V^2+1)\al\be}\,
\right) \; \Phi(\al,\be;k_\rho)\end{array}
$\hfill} (3')

\

where $(\al,\be)$, which replace the parameters $(\om, k_z)$, are

\

\hfill{$
\al \equ \dis{{1 \over {2V}}(\om + V k_z)} \, ; \ \ \ \ \ \be \equ
\dis{{1 \over {2V}}(\om - Vk_z)} \ ,
$\hfill} (5)

\

in terms of the new (``$V$-cone") variables:

\

\hfill{$
\left\{\begin{array}{clr}
\zeta \equ z-Vt  \\
\eta \equ z+Vt \end{array} \right.
$\hfill} (6)

\

The base functions $\psi(\rho,z,t)$ are to be therefore rewritten as

$$ \psi(\rho,\zeta,\eta) \ug J_0(k_\rho \rho) \; \exp{i[\al\ze - \be\eta]} \ .$$

\h Let us now derive the classic ``X-shaped solution", together with some
new[50] generalizations of it.

\h Let us start by choosing the spectrum [with $a > 0$]:

\

\hfill{$
\Phi(\al,\be) \ug \delta(\be - \be') \; \e^{-a\al} \ ,
$\hfill} (7)

\

$a>0$ and $\be'\geq 0$  being constants (related to the transverse and
longitudinal localization of the pulse).

\h In the simple case when $\be'=0$, eq.(3') can be easily integrated over
$\be$ and $k_\rho$ by having recourse to identity (6.611.1) of ref.[51],
yielding

\

\hfill{$
\begin{array}{clr} X \: \equiv \; \Psi_{\Xrm}(\rho,\ze) \ug \dis{\int_{0}^{\infi}\drm \al \,
J_0(\rho\al\sqrt{V^2-1}) \; \e^{-\al(a-i\ze)}} \ = \\ \; \\

\;\;\;\;\;\;\, \ug \left[(a-i\ze)^2 + \rho^2(V^2-1)\right]^{-1/2} \ , \end{array}
$\hfill} (8)

\begin{figure}[!h]
\begin{center}
 \scalebox{.5}{\includegraphics{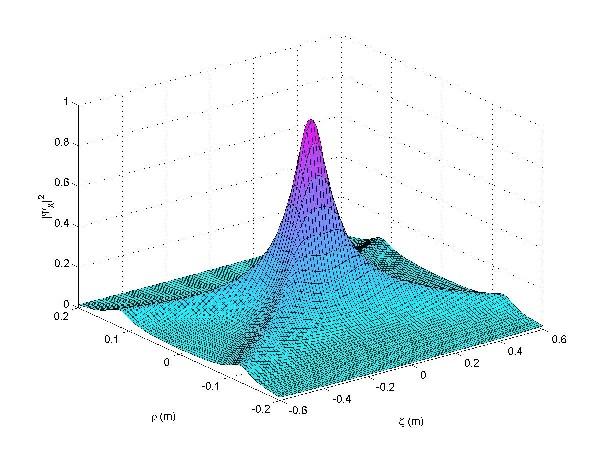}}
\end{center}
\caption{In Fig.10 it is represented (in arbitrary units) the
square magnitude of the ``classic", $X$-shaped Superluminal
Localized Solution (SLS) to the wave equation[41,42], with $V=5c$
and $a=0.1 \;$m: cf. eqs.(8) and (6). \ An infinite family of SLSs
however exists, which generalize the classic $X$-shaped
solution; the Fig.11, below, depicts the first of them (its first
derivative) with the same parameters: see the text and eq.(10).
The successive solutions in such a family are more and more
localized around their vertex.  Quantity $\rho$ is the distance in
meters from the propagation axis $z$, while quantity $\ze$ is the
``$V$-cone" variable (still in meters) $\ze \equiv z-Vt$, with $V
\geq c$. \ Since all these solutions depend on $z$ (and $t$) only
via the variable $\ze$, they propagate ``rigidly", i.e., without
distortion (and are called ``localized", or non-dispersive, for
such a reason).  Here we are assuming propagation in the vacuum
(or in a homogeneous medium).}
\label{fig10}
\end{figure}

\

which is exactly the classic X-shaped solution proposed by Lu \&
Greenleaf[39] in acoustics, and later on by others[42] in electromagnetism,
once relations (4) are taken into account. See Fig.\ref{fig10}.  \ It may be useful to
recall the obvious circumstance that any solution that depends on $z$ (and $t$)
only through the variable $\zeta \equ z-Vt$ will be rigidly moving with
speed $V$.

\h Many other SLSs can be easily constructed; for instance, by inserting
into the weight function (7) the extra factor $\al^m$, \ namely
$\Phi(\al,\be) = \al^m \, \delta(\be) \, \exp[-a\al]$, \ while it is still
$\be'=0$.  Then an infinite family of new SLSs is
obtained (for $m \geq 0$), by using this time identity (6.621.4) of the
same ref.[51]:

\

\hfill{$
\Psi_{\Xrm,m}(\rho,\ze) \ug (-i)^{m}\dis{\frac{\drm^m}{\drm \ze^m}}
\left[(a-i\ze)^2 + \rho^2(V^2-1) \right]^{-1/2}
$\hfill} (9)

\

which generalize[50] the classic X-shaped solution, corresponding to
$m=0$: namely, $\Psi_{\Xrm} \equiv \Psi_{\Xrm,0}$. Notice that all the
derivatives of the latter with respect to $\ze$ lead to new SLSs, all of them
being X-shaped.

\h In the particular case $m=1$, one gets the SLS

\

\hfill{$
\Psi_{\Xrm,1}(\rho,\ze) \ug \dis{\frac{-i \, (a-i\ze)}{\left[(a-i\ze)^2 +
\rho^2(V^2-1)\right]^{3/2}}}
$\hfill} (10)

\begin{figure}[!h]
\begin{center}
 \scalebox{.5}{\includegraphics{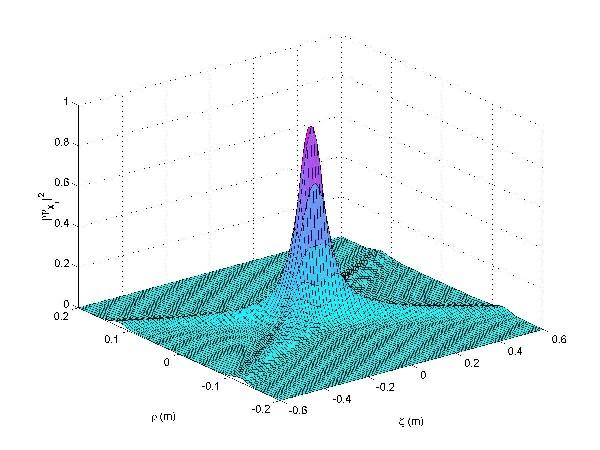}}
\end{center}
\caption{See the caption of the previous Figure 10.}
\label{fig11}
\end{figure}

\

which is the first derivative of the X-shaped wave, and is depicted
in Fig.\ref{fig11}.  One should notice that, by increasing $m$, the pulse becames
more and more localized around its vertex. All such pulses travel, however,
without deforming.

\h Solution (8) is suited for low frequencies only, since its
frequency spectrum (exponentially decreasing) starts from zero. \ One can see
this for instance by writing eq.(7) in the ($\om,k_z$) plane: By eqs.(5) one
obtains

\

\

$$\Phi(\om,k_z) = \delta\left(\frac{\om-Vk_z}{2V}-\be'\right) \;
\dis{\exp[-a \, {{\om+Vk_z} \over {2V}}]}$$

and can observe that $\be'=0$ in the delta implies $\om=Vk_z$. So that the
spectrum becomes $\Phi = \exp[-a\om / V]$, which starts from zero and has a
width given by $\Delta\om=V/a$.

\h By contrast[50], when the factor $\al^m$
is present, the frequency spectrum of the solutions can be ``bumped" in
correspondence with any value $\om_{\Mrm}$ of the angular frequency, provided
that $m$ is large [or $a/V$ is small]: in fact, $\om_{\Mrm}$ results to be
$\om_{\Mrm} = mV / a$. \ The spectrum, then, is shifted towards higher
frequencies (and decays only beyond the value $\om_{\Mrm}$).\\

\

{\bf 3.2} -- {\bf An example of {\rm finite} total-energy solution}\\

\h The previous solutions travel without any deformation, but
possess infinite total-energy (as well as the plane waves).  The
finite total-energy solutions, by contrast, do (slightly, in
general) get deformed while propagating. To show this by an
example, let us go on to a more general choice for the
weight-function: namely, by inserting into eq.(3') for $\be \geq
\be_0$ the spectrum

\

\hfill{$
\Phi \ug \e^{-a\al} \; \e^{-b(\be - \be_0)} \ \ \ \ \ \ \ \ \ \
{\rm for} \ \be \geq \be_0 \ .
$\hfill} (7')

\

We then obtain [for $\be_0 \ll \ 1$] the Superluminal
``Modified Power Spectrum Pulse"[52]

\

\hfill{$
\Psi_{\rm SMPS}(\rho,\ze,\eta) \ug \dis{\e^{b\be_0} \; X \; \int_{\be_0}^\infi
\drm \be \; \e^{-(b+i\eta-Y)\be} \ug X \; \frac{\exp[(Y-i\eta)\be_0]}
{b-(Y-i\eta)}}
$\hfill} (11)

\

in which the integration over $\be$ runs {\em now} from $\be_0$ (no longer
from zero) to infinity; and

\

\hfill{$
Y \ \equiv \ \dis{\frac{V^2+1}{V^2-1} \; \left((a-i\ze)-X^{-1} \right)} \ .
$\hfill}

\

Our SMPS pulses, solutions (11), do not propagate rigidly any
longer (as can be inferred also from the circumstance that they
depend on $\eta$, besides on $\ze$). They are still Superluminal,
however, as can be expected from the fact that they are
superpositions of beams with constant $\theta$ [and therefore with
constant $V$: cf. eqs.(4)]. Actually, on integrating eq.(11) with
$\rho=0$, one gets \ $\Psi_{\rm SMPS}(\rho=0,\ze,\eta) =
\e^{-i\be_0\eta}  \left[(a-i\ze) (b+i\eta)\right]^{-1}$, \ and it
is possible to verify, e.g., that the maximum amplitude of the
(easily evaluated) real part of $\Psi_{\rm SMPS}(0,\ze,\eta)$ goes
on corresponding to $z=Vt$.

\begin{figure}[!h]
\begin{center}
 \scalebox{.5}{\includegraphics{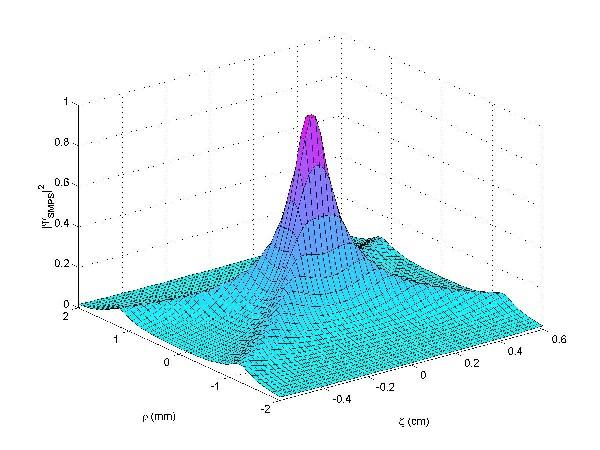}}
\end{center}
\caption{Representation of our Superluminal Modified Power Spectrum (SMPS)
pulses, eq.(11). These beams possess {\em finite} total energy, and
therefore get deformed while traveling. \ This Figure12 depicts the shape of the
pulse, for $t=0$, with $V=5c$, $a=0.001 \;$m, $b=100 \;$m, and $\beta_0 =
1/(100 \, \mrm)$. \ In Fig.13, below, it is shown the same pulse after a 50 meters
propagation.}
\label{fig12}
\end{figure}

\begin{figure}[!h]
\begin{center}
 \scalebox{.5}{\includegraphics{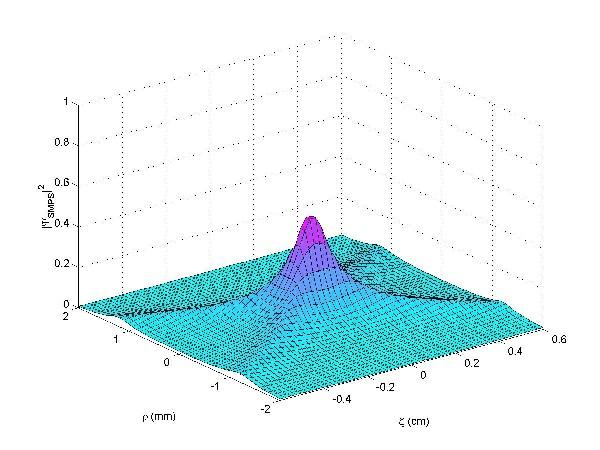}}
\end{center}
\caption{See the caption of the previous Figure 12.}
\label{fig13}
\end{figure}

\h It is worthwhile to emphasize that our solutions (11) possess a
{\em finite total energy}.\footnote{One should recall that the
first finite energy solution, the MFXW, different from but
analogous to our one, appeared in ref.[53].}  \ This is easily
verified. \ In particular, their amplitude decreases with time:
see Fig.\ref{fig12} and Fig.\ref{fig13}.

\h We shall show elsewhere[45] how to construct, e.g., the finite-energy
Superluminal generalization (SSP) of the so-called (luminal) ``splash pulse"
(SP) which was introduced in refs.[52]. \ Our SSP will appear to be the
finite-energy version of the classic X-shaped solution.\\

\

{\large\bf 4. New localized Superluminal exact solutions to the
Maxwell equations, with arbitrary frequencies and adjustable
bandwidths}.\hfill\break

\h Let us finally come to the main topic (of the second part) of this paper. \
>From eq.(2), when putting $c k_{z} \equiv n \om \, \cos\theta$, one
can write $c k_{\rho} \equiv n \om \, \sin\theta$. \ In the new
coordinates ($\om$,$\theta$), solution (1) becomes

\

\hfill{$ \dis{\psi(\rho,\ze) \ug J_0(\frac{n \om}{c} \rho \,
\sin\theta) \; \exp\left[+i \frac{n \om}{c}\ze \cos\theta \right]}
$\hfill} (12)

\

where we used the relations $\ze \equiv z - Vt$ and $V = c/(n \,
\cos\theta)$, and we confine ourselves for the moment to a
homogeneous medium with $n \; = \;$constant. \ Equation (12)
contains two free parameters: $\om$ and $\theta$. \ This equation
(which represents the ``Bessel beam") tells us once more that such
a beam is transversally localized in energy, and propagates
without dispersion.  Considering a superposition of waves with
different frequencies and $\theta$ (and therefore $V$)
constant,\footnote{Superpositions of beams with different values
of $\theta$ (and therefore of $V$) will be considered elsewhere.}
one can get the SLSs in the alternative form

\

\hfill{$ \dis{\Psi (\rho,\ze) \ug \int_{0}^{\infty} S(\om)
J_0(\frac{n \om}{c} \rho \, \sin\theta) \; \exp\left[+i \frac{n
\om}{c}\ze \cos\theta\right]} \drm\om \ . $\hfill} (13)

\

We have to find out a spectrum $S(\om)$ which preserves the integrability of
eq.(13) for any frequency range: In order to be able to shift our spectrum
towards the desired frequency, let us locate it around a central frequency,
$\om_\crm$, with an (arbitrary) bandwidth $\Delta\om$.\ Then, let us choose
the spectrum

\

\hfill{$
S(\om) \ug \dis{\left(\frac{\om}{V}\right)^m\e^{-a\om}}
$\hfill} (14)

\

where $V$ is the wave velocity, while $m$ and $a$ are free
parameters. [For $m=0$, it is $S(\om)=\exp[-a\om]$, and one
gets the (standard) X-wave spectrum]. \ For $m \neq 0$, some mathematical
manipulations yield the relations

\

\hfill{$
\dis{m^{-1} \ug (\Delta\om_{\pm}/\om_\crm) - \ln [1+ (\Delta\om_\pm /
\om_\crm)]}
$\hfill} (14a)

\

\hfill{$
\dis{\om_\crm = \frac{m}{a}} \ .
$\hfill} (14b)

\begin{figure}[!h]
\begin{center}
 \scalebox{.5}{\includegraphics{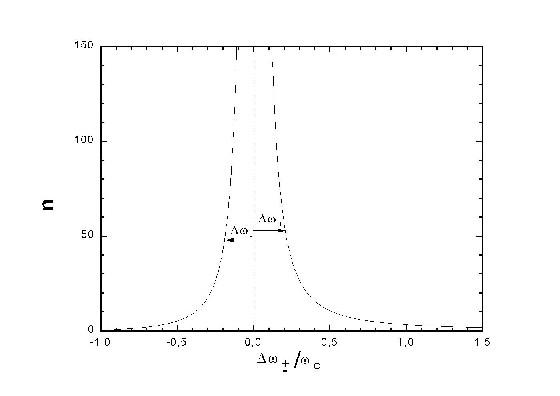}}
\end{center}
\caption{Behavior of the derivative number, $m$, as a function of
the normalized bandwidth frequency, $\Delta\om_{\pm}/\om_\crm$.
Given a central frequency, $\om_\crm$, and a bandwidth,
$\Delta\om_{\pm}$, one finds the exact value of $m$ by
substituting these values into eq.(14a).}
\label{fig14}
\end{figure}

\

where, because of the non-symmetric character of spectrum (14), we had to
call  $\Delta\om_+$ ($>0$) and $\Delta\om_-$ ($<0$) the
bandwidth to the right and to the left of $\om_\crm$, respectively; so
that $\Delta\om = \Delta\om_+ - \Delta\om_- $. [However, already for
small values of $m$ (typically, for $m\geq 10$), one
has $\Delta\om_+ \approx - \Delta\om_-$]. \ Once defined
$\om_\crm$ and $\Delta\om$, one can determine $m$ from the
first equation. Then, using the second one, $a$ is found. \
Figure \ref{fig14} illustrates the behavior of relation (14a): One can observe that
the smaller $\Delta\om / \om_\crm$ is, the higher $m$ must be; parameter
$m$ plays the important role of controlling the spectrum bandwidth.

\begin{figure}[!h]
\begin{center}
 \scalebox{.5}{\includegraphics{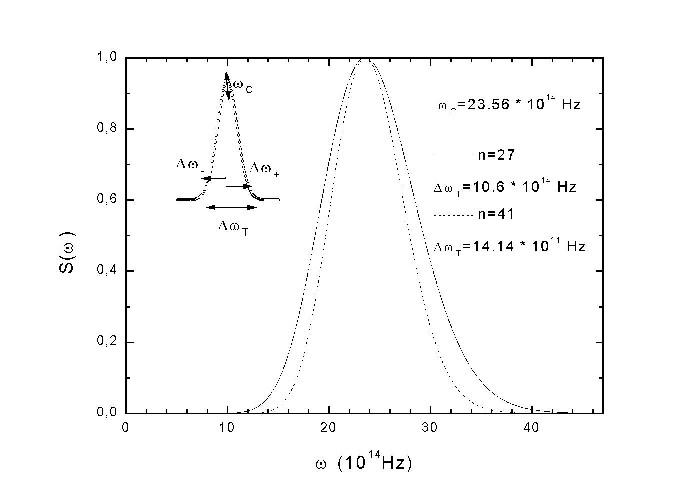}}
\end{center}
\caption{Normalized spectra for $\om_\crm=23.56 \times 10^{14} \;$Hz
and different bandwidths. The solid line corresponds to $m=27$, and the
dotted line to $m=41$. See the text. \
Both spectra have {\em the same} $\om_\crm$: To get this result, one can
observe that (taking the narrower spectrum as a reference)
both $m$ and $a$ have to increase. This figure shows moreover the important
role of $m$ in controlling the spectrum bandwidth.}
\label{fig15}
\end{figure}

\h From the X-wave spectrum, it is known that $a$ is related to
the (negative) slope of the spectrum. \ On the contrary,
quantity $m$ has the effect of rising the spectrum. In this way,
one parameter compensates for the other, producing the
localization of the spectrum inside a certain frequency range. At
the same time, this fact also explains (because of relation
(14b)) why an increase of both $m$ and $a$ is necessary to keep
the same $\om_\crm$. This can be seen from Fig.\ref{fig15}.

\h One can verify that, on inserting into eq.(13) the exponential spectrum
$S(\om) = \exp[-a\om]$, one gets once more the ordinary X-wave, namely

$$X \ug V \left[\sqrt{(aV-i\ze)^2+\rho^2({n_0}^2 V^2/c^2-1)}\right]^{-1} \; ,$$

where now we
explicitly wrote down the refraction index $n_\orm$ of the considered
(dispersionless) medium. \ To exemplify the use of our solution
(13), let us rewrite it with the spectrum (14):

\

\hfill{$
\Psi(\rho,\ze) \ug \dis{
V\,\int_{0}^{\infty}}(\frac{\om}{V})^m J_0 \left( \frac{\om}{V} \rho
\sqrt{n_0^2\frac{V^2}{c^2}-1} \right) \; \dis{\e^{-(aV-i\ze)\om/V}}
\drm(\frac{\om}{V})
$\hfill} (15)

\

which shows that the use of a spectrum like (14) allows shifting the
proposed solution towards any frequency while confining it within the desired
frequency range. \

\h It is not difficult to show (or verify) that eq.(15) can be written in
the interesting forms

\

\hfill{$
\Psi(\rho,\ze) \ug \dis{(-1)^m \frac{\partial^m\,X}{\partial(aV-i\ze)^m}} \ ,
$\hfill} (16)

\

and, by use of identity (6.621) of ref.[51],

\

\hfill{$
\Psi(\rho,\ze) \ug
\dis{\frac{\Gamma(m+1)\,X^{m+1}}{V^{m}} \, \, F\left(\frac{m+1}{2}\, ;
\, -\frac{m}{2} \, ; \, 1 \, ; \, (n_0^2\frac{V^2}{c^2}-1)\,\rho^2\,
\frac{X^2}{V^2} \right)} \ ,
$\hfill} (16')

\

where $F$ is the Gauss hypergeometric function. [Equation (16') can be
useful in the cases of large values of $m$].

\begin{figure}[!h]
\begin{center}
 \scalebox{.55}{\includegraphics{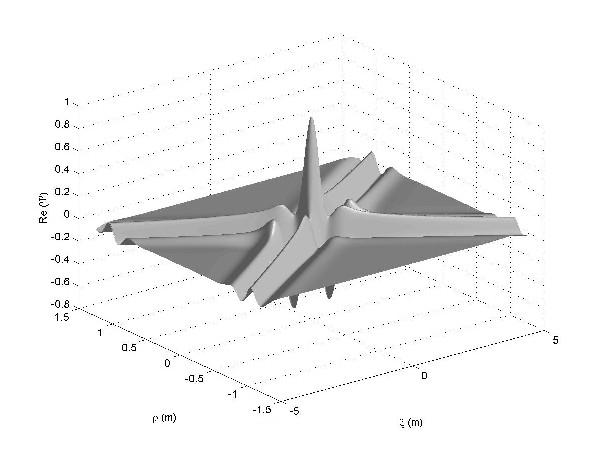}}
\end{center}
\caption{The real part of an X-shaped beam for microwave frequencies
in a dispersionless medium. It corresponds to $\om_\crm = 6 \times 10^9 \;$
GHz and $\Delta\om = 0.9 \, \om_\crm$, while the values of $m$
and $a$, calculated by using eqs.(14a)) and (14b), are $m=10$ and
$a=1.6667 \times 10^{-9}$. The resulting wave possesses both a longitudinal
and a transverse localization.}
\label{fig16}
\end{figure}

\h Let us call attention to equations (16) and (16'): to our
knowledge,\footnote{It can be noticed that $\partial X/\partial
(aV-i\ze)= (iV)^{-1} \partial X/\partial t$. Time derivatives of
the X-wave have been actually considered by J.Fagerholm et
al.[50]: however the properties of the generating spectrum (like
its ability in frequency shifting and bandwidth controlling) have
not been studied in previous work.} no analytical expression has
been previously met for X-{\em type} waves, apt at being localized
in the neighborhood of any chosen frequency with an adjustable
bandwidth. \ Figure \ref{fig16} shows an example of X-type wave
for microwave frequencies: it has shape and properties similar
to the classic X-wave's.\\

\

{\large\bf 5. Superluminal localized waves in dispersive media}.\hfill\break

\h Let us now go on to dealing with {\em dispersive} media, in which case
the refraction index depends on the wave frequency: $n = n(\om)$; so that
we shall have[49]

\

\hfill{$
\left\{
\begin{array}{l}
c k_{z} \equiv n(\om) \, \om \, \cos\theta \\
\\
c k_{\rho} \equiv n(\om) \, \om \, \sin\theta \; .
\end{array}
\right.
$\hfill}

\

Looking for a localized wave that does not suffer dispersion, one
can exploit the fact that $\theta$ determines the wave velocity:
Namely, one can choose a particular frequency dependence of
$\theta$ to {\em compensate} for the (geometrical) dispersion due
to the variation of $n$ with the frequency.[49] \ When a
dispersionless pulse is desired, the constraint $k_{z} = d + \om
\, b$ must be satisfied. On using the first one of the last two
equations, we infer such a constraint to be forwarded by the
following relationship between $\theta$ and $\om$:

\

\hfill{$ \cos(\theta(\om))=\dis{\frac{(d + b\,\om)c}{\om\;
n(\om)}} \ , $\hfill} (17)

\

with $d$ and $b$ arbitrary constants ($b$ being related to the wave velocity:
$b=1/V$). \ If we choose, for convenience, $d=0$, eq.(13) becomes

\

\hfill{$
\Psi (\rho,\ze) \ug \dis{
\int_{0}^{\infty}}S(\om)\,J_0 \left(\rho\om
\sqrt{\frac{n^2(\om)}{c^2}-b^2}\,\right) \; \dis{\e^{+i\om
b\ze}}\, \drm\om \ .
$\hfill} (18)

\

\h Let us stress that eq.(18) constitutes an analytical (integral) formula
which
represents a wave propagating without dispersion in a {\em dispersive}
medium; it is well suited for any desired frequency range, and a priori
can find applications in acoustics, microwave physics, optics, etc.

\begin{figure}[!h]
\begin{center}
 \scalebox{.5}{\includegraphics{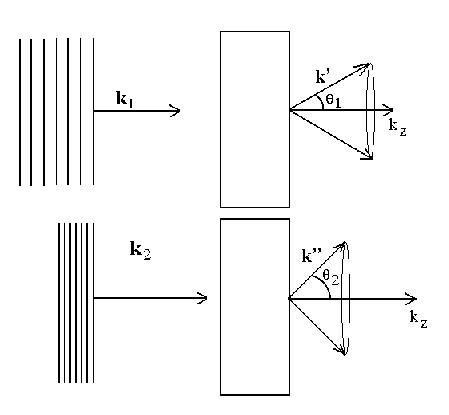}}
\end{center}
\caption{Sketch of a generic device (axicone, hologram, etc.)
suited to properly deviating the wave vector of each spectral
component.}
\label{fig17}
\end{figure}

\h Let us mention how it is possible to realize relation (17) for
{\em optical} frequencies, by following a procedure similar to the
one illustrated in Figure \ref{fig17}. The wave vector for each
spectral component suffers a different deviation in passing
through the chosen device (axicone, hologram, and so
on[41-43,54]): and such a deviation, associated with the
dispersion due to the medium, can be such that the phase velocity
remains the same for every frequency. This corresponds to no
dispersion for the beam. \ We may consider a nearly gaussian
spectrum, like that in eq.(14), and assume the refractive index
(for the frequency range of interest) to depend linearly on $\om$,
that is, $n(\om) = n_0 + \om \, \delta$, with $\delta$ a free
parameter: a linear dependence that is actually realizable for
frequencies far from the resonances. \ In this case, on using
spectrum (14), a relation similar to eq.(22) is found, which can
be evaluated via a Taylor expansion, that (if $\delta$ is small
enough) can be truncated at its first derivative.  The expression
of such a first derivative can be integrated (by using identity
6.621.4 of ref.[51]); and eventually we find that our eq.(18)
admits the approximate solution

\

\hfill{$
\Psi(\rho,\ze) =
(-1)^m\dis{\frac{\partial^m\,X}{\partial(aV-i\ze)^m}} +
(-1)^{m+4}\frac{V^3\,n_0}{c^2\,\left(n_0^2\frac{V^2}{c^2}-1\right)}
\dis{\frac{\partial^{m+2}}{\partial(aV-i\ze)^{m+2}}}
\left[(aV-i\ze)\,X\right] \, \delta
$\hfill} (19)

\

which could be written also in a way similar to eq.(16'). \ It is
interesting to notice that also the SLS, eq.(19), for a dispersive medium
has been obtained by simple mathematical operations (derivatives) acting
to the standard ``X-wave''.\\

\

{\large\bf 6. -- Optical applications}\\

\h Let us make two practical examples, both for optical frequencies: in
particular, with optical fibers. The bulk of the dispersive medium
be fused Silica ($\SiO_2$). Far from the medium resonances (which is our case),
the refractive index can be approximated by the well-known Sellmeier
equation[55]

\

\hfill{$
n^2(\om) \ug \dis{1\,+\,\sum_{j=1}^{N}\,\frac{B_j\,\om_j^2}{\om_j^2-
\om^2}} \ ,
$\hfill} (20)

\begin{figure}[!h]
\begin{center}
 \scalebox{1.25}{\includegraphics{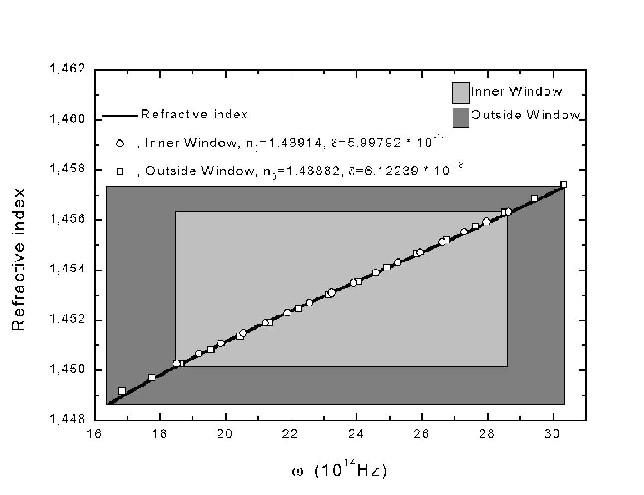}}
\end{center}
\caption{Variation of the refractive index $n(\om)$ with
frequency for fused Silica. The solid line is its behaviour,
according to Sellmeir's formulae. The open circles and squares
are the linear approximations for $m=41$ and $m=27$,
respectively. See the text.}
\label{fig18}
\end{figure}

\

where $\om_j$ are the resonance frequencies, $B_j$ the strength
of the $j$th resonance, and $N$ the total number of the
material resonances that appear in the frequency range of
interest. For typical frequencies of ``long-haul transmission" in
optics, it is appropriate to choose $N=3$, which yields the
values[55] $B_1=0.6961663$; \ $B_2=0.4079426$; \ $B_3=0.8974794$; \
$\lambda_1=0.0684043 \; \mu$m; \ $\lambda_2=0.1162414 \; \mu$m; \
and $\lambda_3=9.896161 \; \mu$m. \ Figure \ref{fig18} illustrates the relation
between $n$ and $\om$, and specifies the range we are going to adopt here.
Our two examples consist in localizing the spectrum around the angular
frequency $\om_c=23.56 \times 10^{14} \;$Hz (which corresponds to the
wavelength $\la_c=0.8\; \mu$m, with the two different bandwidth
$\Delta\om_1 = 0.55 \om_c$ and $\Delta\om_2 = 0.4 \om_c$. The
values of $a$ and $m$ corresponding to these two situations are
$a=1.14592 \times 10^{-14}$, $m=27$, and $a = 1.90986 \times
10^{-14}$, $m=41$, respectively. \ On looking at the two ``windows'' in
Fig.18, one notices that the Silica refraction index does
not suffer strong variations, and that a linear approximation to $n=n(\om)$
is quite satisfactory. Moreover, for both the situations and
their respective $n_0$ values, the value of $\delta$ results to be very
small (which means that in these cases the mentioned truncation of the
Taylor expansion is certainly acceptable).

\begin{figure}[!h]
\begin{center}
 \scalebox{.5}{\includegraphics{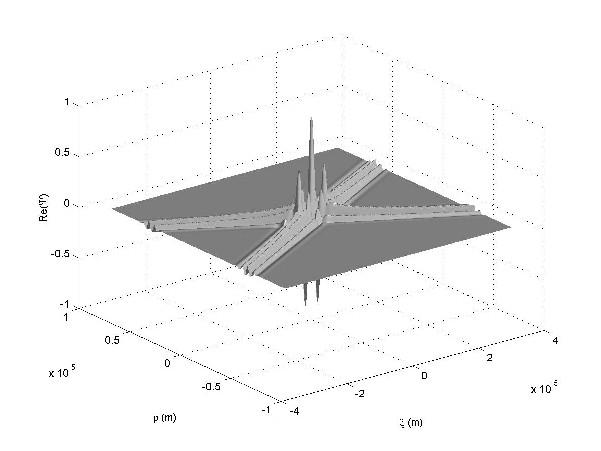}}
\end{center}
\caption{The real part of an X-shaped beam for optical
frequencies {\em in a dispersive medium}, with $m=27$. It refers to the
larger window in Fig.16.}
\label{fig19}
\end{figure}

\begin{figure}[!h]
\begin{center}
 \scalebox{.5}{\includegraphics{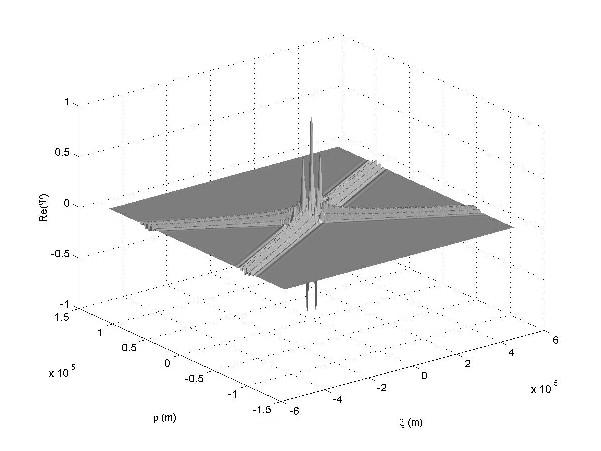}}
\end{center}
\caption{The real part of an X-shaped beam for optical
frequencies {\em in a dispersive medium}, with $m=41$.  It refers to the
inner window in Fig.16.}
\label{fig20}
\end{figure}

\h The beam intensity profiles for the two bandwidths are shown in
Fig.\ref{fig19} and Fig.\ref{fig20}. \ In the first figure, one
can see a pattern similar to that of an ordinary X-wave (cf.
Fig.3a); but now the pulse is much more localized spatially and
temporally (typically, it is a femtosecond pulse). \ The second
figure shows some small differences, with respect to the first
one, in the spatial oscillations inside the wave {\em
envelope\/}[56]: This

can be due to the circumstance that the carrier wavelength, for
certain values of the bandwidth, becomes shorter than the width of
the spatial envelope; so that one meets a well defined carrier
frequency. \ Let us repeat that both these waves are transversally
{\em and} longitudinally localized, being moreover free from
dispersion (since the dependence of $\Psi$ on $z$ (and $t$) is
through the variable $\ze =z-Vt$ only) just as it happens for a
classic X-shaped wave.\\

{\large\bf 7. -- Conclusions}\\

\h In the first part of this article (after a brief theoretical
introduction based on Special Relativity) we have mentioned the
various experimental sectors of physics in which Superluminal motions
seem to appear. In particular, a bird's-eye view has been
presented of the experiments with evanescent waves (and/or tunneling
photons), and  with the ``localized Superluminal solutions" (SLS) to the wave
equation, like the so-called X-shaped beams; we have added to such an
``experimental", very short review a number of references, sufficient in some
cases to provide the interested readers with reasonable bibliographical
information.

\h In the second part of this paper, we have first worked out analytical
Superluminal localized solutions to the wave equation for arbitrary
frequencies and with adjustable bandwidth in vacuum. The same methodology has
been then used to obtain new, analytical expressions representing
X-type waves (with arbitrary frequencies and adjustable
bandwidth) which propagate {\em in dispersive media}. Such expressions
have been obtained, on one hand, by adopting the appropriate spectrum
(which made possible to us both choosing the carrier frequency rather
freely, and  controlling the spectral bandwidth), and,
on the other hand, by having recourse to simple mathematics.
Finally, we have illustrated some examples of our approach with
applications in optics, considering fused Silica as the
dispersive medium.

\h The case of the very interesting
(and more orthodox, in a sense) {\em subluminal} Localized Waves,
solutions to the wave equations, will be dealt with in another paper.\\

{\bf Acknowledgements}\\

This paper (partly a review) was prepared for the special issue on ``Nontraditional
Forms of Light" of the
IEEE JSTQE: one of the authors (ER) is grateful to all the Editors of that issue, and
in particular to P.W.Milonni and H.Winful. The initial part of this paper
was written while he was visiting the KITP of the UCSB within the activities
of a Workshop on quantum optics: and he wishes to thank the Organizers (in
particular Ray Chiao and Peter Milonni) for their kind invitation, as
well as David Gross and Daniel Hone for generous hospitality. All the authors
are moreover grateful for useful discussions or kind cooperation, to G.Battistoni,
M.Brambilla, C.Cocca, C.Conti, J.H.Eberly, G.Degli Antoni, M.Fleishhauer, F.Fontana,
M.A.Porras, S.Martellucci, M.Mattiuzzi, P.Pizzochero, P.Saari, G.Salesi and A.Shaarawi.
Finally, the authors gratefully acknowledge
a grant from Alberto Castoldi, the Rector of the Bergamo state University,
which allowed paying the publication charges of this paper. \ This article
first appeared, in preliminary form, as preprint NSF-ITP-02-93
(I.T.P., UCSB; 2002).\\

\newpage

{\bf Bibliography:}\hfill\break

[1] See, e.g., O.M.Bilaniuk, V.K.Deshpande and E.C.G.Sudarshan,
``Meta relativity", {\em Am. J. Phys.}, vol.30,
p.718, Oct.1962. \ Cf. also T.Alv\"ager and M.N.Kreisler, ``Quest for
faster-than-light particles", {\em Phys. Rev.}, vol.171,
pp.1357-1361, Feb.1968.\hfill\break

[2] See E.Recami and R.Mignani, ``Classical Theory of Tachyons
(Special Relativity Extended to Superluminal Frames and
Objects)", {\em Rivista N. Cim.}, vol.4, pp.209-290, 1974, and
refs. therein. \ Cf. also E.Recami (editor), {\em Tachyons, Monopoles, and
Related Topics} (North-Holland; Amsterdam, 1978); \ H.C.Corben,
``Tachyon matter and complex physical variables", {\em Nuovo
Cimento A}, vol.29, pp.415-426, 1975.\hfill\break

[3] E.Recami, ``Classical tachyons and possible applications",
{\em Rivista N. Cim.}, vol.9(6), pp.1-178, 1986, issue no.6,
and refs. therein.\hfill\break

[4] See, e.g., E.Recami, ``Superluminal motions? A bird's-eye
view of the experimental situation", {\em Found. Phys.}, vol.31,
pp.1119-1135, July 2001; \ ``I Tachioni", in {\em Annuario '73,
Enciclopedia EST}, E.Macorini editor (Mondadori; Milan, 1973), pp.85-94; \
and \ ``I tachioni", {\em Nuovo Saggiatore}, vol.2(3), pp.20-29, 1986,
issue no.3.\hfill\break

[5] E.Recami, ``Special relativity and causality" in {\em
I Concetti della Fisica}, F.Pollini and G.Tarozzi editors (Acc. Naz.
Sc. Lett. Arti; Modena, 1993), pp.125-138; \ E.Recami and W.A.Rodrigues,
``Antiparticles from Special Relativity with
ortho-chronous and anti-chronous Lorentz transformations", {\em
Found. Phys.}, vol.12, pp.709-718, July 1982; vol.13, E533 (1983).\hfill\break

[6] E.Recami, ``Tachyon mechanics and causality; A systematic
thorough analysis of the tachyon causal paradoxes", {\em Found.
Phys.}, vol.17, pp.239-296, May 1987. \ See also E.Recami, ``The
Tolman-Regge antitelephone paradox: Its solution by tachyon
mechanics", {\em Lett. Nuovo Cimento}, vol.44, pp.587-593, Nov.1985; \
P.Caldirola and E.Recami, ``Causality and tachyons in
Relativity", in {\em Italian Studies in the Philosophy of Science},
M.Dalla Chiara editor (Reidel; Boston, 1980), pp.249-298; \ A.M.Shaarawi
and I.M.Besieris, ``Relativistic causality and superluminal signalling using
X-shaped localized waves", {\em J. Phys., A}, vol.33, pp.7255-7263,
Oct.2000.\hfill\break

[7] Cf. M.Baldo Ceolin, ``Review of neutrino physics", invited talk at the
{\em XXIII Int. Symp. on Multiparticle Dynamics (Aspen, CO; Sept.1993)}; \
E.W.Otten, ``Direct neutrino mass measurements", {\em Nucl.
Phys. News}, vol.5, pp.26-35, Jan.1995. \ From the theoretical
point of view, see, e.g., E.Giannetto, G.D.Maccarrone,
R.Mignani and E.Recami, ``Are muon neutrinos faster-than-light
particles? ", {\em Phys. Lett., B}, vol.178, pp.115-120, Sept.1986,
and refs. therein; \ S.Giani, ``Experimental evidence of
superluminal velocities in astrophysics and proposed
experiments", CP458, in {\em Space Technology and Applications International
Forum 1999}, ed. by M.S.El-Genk (A.I.P.; Melville, 1999), pp.881-888.\hfill\break

[8] See, e.g., J.A.Zensus and T.J.Pearson (editors), ``Superluminal
Radio Sources`` (Cambridge University Press; Cambridge, 1987).\hfill\break

[9] I.F.Mirabel and L.F.Rodriguez, ``A superluminal source in the Galaxy",
{\em Nature}, vol.371, pp.46-48, Sept.1994 [with an editorial comment,
``A galactic speed record", by G.Gisler, at page 18 of the same issue]; \
S.J.Tingay et al., ``Relativistic motion in a nearby bright X-ray source",
{\em Nature}, vol.374, pp.141-143, March 1995.\hfill\break

[10] M.J.Rees, ``Appearance of relativistically expanding radio
sources", {\em Nature}, vol.211, pp.468-471, Jan.1966; \ A.Cavaliere,
P.Morrison and L.Sartori, ``Rapidly changing radio images",
{\em Science}, vol.173, pp.525-529, May 1971.\hfill\break

[11] E.Recami, A.Castellino, G.D.Maccarrone and M.Rodon\`o,
``Considerations about the apparent Superluminal expansions
observed in astrophysics", {\em Nuovo Cimento B}, vol.93,
pp.119-153, June 1986. \ Cf. also R.Mignani and E.Recami,
``Possibility of Superluminal Sources and their Doppler Effect",
{\em Gen. Relat. Grav.}, vol.5, pp.615-620, May 1974.\hfill\break

[12] V.S.Olkhovsky and E.Recami, ``Recent developments in the
time analysis of tunneling processes", {\em Phys. Reports},
vol.214, pp.339-356, May 1992, and refs. therein: In particular,
T.E.Hartman, ``Tunneling of a wave packet", {\em J. Appl. Phys.},
vol.33, pp.3427-3433, Dec.1962; \ L.A.MacColl, ``Note on the
transmission and reflection of wave packets by potential
barriers", {\em Phys. Rev.}, vol.40, pp.621-626, May 1932. \ See
also V.S.Olkhovsky, E.Recami, F.Raciti and A.K.Zaichenko,
``More about tunneling times, the dwell time and the Hartman
effect", {\em J. de Physique-I (France)}, vol.5, pp.1351-1365, Oct.1995; \
M. Abolhasani and M.Golshani: ``Tunneling times in the Copenhagen interpretation
of quantum mechanics", {\em Phys. Rev., A}, vol.62, 012106, June 2000 (7
pages); \  V.S.Olkhovsky, E.Recami and J.Jakiel, ``Unified time
analysis of photon and nonrelativistic particle tunneling``, available online
c/o www.arXiv.org as the e-print physics/0102007, Feb.2001 (subm. to {\em
Phys. Reports\/}); \ P.W.Milonni, see ref.[22] below.\hfill\break

[13] See, e.g., R.Y.Chiao, P.G.Kwiat and A.M.Steinberg,
``Analogies between electron and photon tunneling: A proposed
experiment to measure photon tunneling times", {\em Physica B},
vol.175, pp.257-262, Dec.1991; \ A.Ranfagni, D.Mugnai,
P.Fabeni and G.P.Pazzi, ``Delay-time measurements in narrowed
wave-guides as a test of tunneling", {\em Appl. Phys. Lett.},
vol.58, pp.774-776, Feb.1991; \ Th.Martin and R.Landauer,
``Time delay of evanescent electromagnetic waves and the analogy
to particle tunneling", {\em Phys. Rev., A}, vol.45,
pp.2611-2617, Feb.1992; \ Y.Japha and G.Kurizki, ``Superluminal
delays of coherent pulses in nondissipative media: A universal
mechanism", {\em Phys. Rev., A}, vol.53, pp.586-590, Jan.1996.\hfill\break

[14] E.Recami, F.Fontana and R.Garavaglia, ``Special
relativity and Superluminal motions: A discussion of some recent
experiments", {\em Int. J. Mod. Phys., A}, vol.15, pp.2793-2812,
July 2000, and refs. therein.\hfill\break

[15] A.M.Steinberg, P.G.Kwiat and R.Y.Chiao, ``Measurement of the single-photon
tunneling time", {\em Phys. Rev. Lett.}, vol.71, pp.708-711,
Aug.1993, and refs. therein; \ R.Y.Chiao, P.G.Kwiat and A.M.Steinberg,
``Faster than light", {\em Scient. Am.}, vol.269(2), pp.52-60, Aug.1993,
issue no.2. \ For a review, see, e.g., R. Y.Chiao and A.M.Steinberg, ``Tunneling
times and superluminality", {\em Progress in Optics}, E.Wolf editor (Elsevier;
Amsterdam, 1997), pp.347-406.\hfill\break

[16] A.Enders and G.Nimtz, ``On superluminal barrier traversal",
{\em J. de Physique-I}, vol.2, pp.1693-1698, Sept.1992; \
``Zero-time tunneling of evanescent mode packets", {\em ibidem},
vol.3, pp.1089-1092, May 1993; \ ``Evanescent-mode propagation and
quantum tunneling", {\em Phys. Rev., E}, vol.48, pp.632-634, July
1993; \ G.Nimtz, A.Enders and H.Spieker, ``Photonic tunneling
times", {\em J. de Physique-I}, vol.4, pp.565-570, Apr.1994; \
H.M.Brodowsky, W.Heitmann and G.Nimtz, see refs.[20] below. \ For a review,
see, e.g., G.Nimtz and W.Heitmann, ``Superluminal
photonic tunneling and quantum electronics", {\em Prog. Quant. Electr.},
vol.21, 81 (1997).\hfill\break

[17] A.Ranfagni, P.Fabeni, G.P.Pazzi and D.Mugnai,
``Anomalous pulse delay in microwave propagation: A plausible
connection to the tunneling time", {\em Phys. Rev., E}, vol.48,
pp.1453-1460, Aug.1993; \ Ch.Spielmann, R.Szipocs, A.Stingl and
F.Krausz, ``Tunneling of Optical Pulses through Photonic Band
Gaps", {\em Phys. Rev. Lett.}, vol.73, pp.2308-2311, Oct.1994; \
Ph.Balcou and L.Dutriaux, ``Dual Optical Tunneling Times in
Frustrated Total Internal Reflection", {\em Phys. Rev. Lett.},
vol.78, pp.851-854, Feb.1997; \  V.Laude and P.Tournois,
``Superluminal asymptotic tunneling times through one-dimensional
photonic bandgaps in quarter-wave-stack dielectric mirros", {\em
J. Opt. Soc. Am., B}, vol.16, pp.194-198, Jan.1999.\hfill\break

[18] Scientific American (Aug.1993); {\em Nature} (Oct.21,
1993); New Scientist (Apr.1995); Newsweek (19 June 1995).\hfill\break

[19] Ref.[3], p.158 and pp.116-117. Cf. also D.Mugnai, A.Ranfagni,
R.Ruggeri, A.Agresti and E.Recami, ``Superluminal
processes and signal velocity in tunneling simulation", {\em
Phys. Lett., A}, vol.209, pp.227-234, Dec.1995.\hfill\break

[20] H.M.Brodowsky, W.Heitmann and G.Nimtz, ``Comparison of
experimental microwave tunneling data with calculations based on
Maxwell's equations", {\em Phys. Lett., A}, vol.222, pp.125-129,
Oct.1996.\hfill\break

[21] A.P.L.Barbero, H.E.Hernandez F., and E.Recami,
``Propagation speed of evanescent modes" [e-print physics/9811001],
{\em Phys. Rev., E}, vol.62, pp.8628-8635, Dec.2000, and refs. therein. \
See also E.Recami, H.E.Hernandez F., and A.P.L.Barbero, ``Superluminal
microwave propagation and special relativity", {\em Ann. der
Physik (Leipzig)}, vol.7, pp.764-773, 1998; \ A.M.Shaarawi and I.M.
Besieris, ``Superluminal tunneling of an electromagnetic X-wave
through a planar slab", {\em Phys. Rev., E}, vol.62, pp.7415-7420,
Nov.2000.\hfill\break

[22] Besides the classic works by A.Sommerfeld, ``Ein Einwand
gegen die Relativtheorie der Elektrodynamik und seine
Beseitigung", {\em Z. Physik}, vol.8, pp.841-842, 1907,
and by L.Brillouin, ``\"Uber die Fortpflanzung des Lichts in
dispergierenden Medien", {\em Ann. der Physik}, vol.44, pp.203,
1914; \ {\em Wave Propagation and Group Velocity}
(Academic Press; New York, 1969), see, e.g., P.W.Milonni,
``Controlling the speed of light pulses", {\em J. Phys., B},
vol.35, R31-R56, March 2002; \ R.W.Ziolkowski, ``Superluminal
transmission of information through an electromagnetic
metamaterial", {\em Phys. Rev., E}, vol.63, 046604,
Apr.2001 (13 pages); \ D.Mugnai, ``Bessel beam and signal propagation",
{\em Phys. Lett., A}, vol.278, pp.6-10, 2000, \ M.Zamboni-Rached
and D.Mugnai, ``Erratum to `Bessel beams and signal propagation',"
{\em Phys. Lett., A}, vol.284, pp.294-295, June 2001; \ E.Recami,
F.Fontana and R.Garavaglia, ``Special relativity and Superluminal
motions: A discussion of some recent experiments", {\em Int. J.
Mod. Phys., A}, vol.15, pp.2793-2812, July 2000, and refs.
therein; \ A.M.Shaarawi and I.M.Besieris, ``On the
superluminal propagation of X-shaped localized waves", {\em J.
Phys., A}, vol.33, pp.7227-7254, Oct.2000; \
G.Nimtz and A.Haibel, ``Basic of superluminal
signals", {Ann. der Physik}, vol.11(2), pp.163-171, 2000; \ G.Nimtz,
``Evanescent modes are not necessarily Einstein causal",
{\em Europ. Phys. Journal B}, vol.7(4), pp.523-525, 1999; \ G.Kurizki,
A.E.Kozhekin and A.G.Kofman, ``Tachyons and information
transfer in quantized parametric amplifiers", {\em Europhys. Lett.}, vol.42,
499-504, 1998; \ G.Kurizki, A.E.Kozhekin, A.G.Kofman and M.Blaauboer,
``Optical tachyons in parametric amplifiers: How fast can quantum information
travel?", {\em Opt. Spektrosc.}, vol.87, 505-512, 1999.\hfill\break

[23] L.J.Wang, A.Kuzmich and A.Dogariu, ``Gain-assisted
superluminal light propagation", {\em Nature}, vol.406,
pp.277-279, July 2000. \ Cf. also, e.g., G.Toraldo di Francia, ``Super-gain
antennas and optical resolving power", {\em Supplem. Nuovo Cim.}, vol.9(3),
pp.391-435, 1952.\hfill\break

[24] G.Nimtz, A.Enders and H.Spieker, in {\em Wave and Particle in
Light and Matter}, A.van der Merwe and A.Garuccio editors (Plenum;
New York, 1993). \ See also A.Enders and G.Nimtz, ``Photonic tunneling
experiments", {\em Phys. Rev., B}, vol.47, pp.9605-9609, May 1993.\hfill\break

[25] S.Longhi, P.Laporta, M.Belmonte and E.Recami,
``Measurement of superluminal optical tunneling times in
double-barrier photonic band gaps", {\em Phys. Rev., E}, vol.65,
046610, Apr.2002 (6 pages).\hfill\break

[26] V.S.Olkhovsky, E.Recami and G.Salesi, ``Superluminal
tunneling through two successive barries" [e-print quant-ph/0002022],
{\em Europhys. Lett.}, vol.57, pp.879-884, Feb.2002 ; \ Y.Aharonov,
N.Erez and B.Reznik, ``Superoscillations and tunneling times", {\em Phys.
Rev., A}, vol.65, 052124, May 2002 (5 pages).\hfill\break

[27] V.S.Olkhovsky, E.Recami and G.Salesi, refs.[26]; \
S.Esposito, ``Multibarrier tunneling", e-print quant-ph/020918. \ See
also A.M.Shaarawi and I.M.Besieris, ``Ultra-fast multiple tunnelling of
electromagnetic X-waves", {\em J. Phys., A}, vol.33, pp.8559-8576,
Dec.2000.\hfill\break

[28] V.S.Olkhovsky, E.Recami, F.Raciti and A.K.Zaichenko, ref.[12],
page 1361 and refs. therein. \ See also refs.[3,6], and E.Recami,
F.Fontana and R.Garavaglia, ref.[14], page 2807 and refs.
therein.\hfill\break

[29] R.Y.Chiao, A.E.Kozhekin, and G.Kurizki, ``Tachyonlike
Excitations in Inverted Two-Level Media", {\em Phys. Rev. Lett.},
vol.77, pp.1254-1257, Aug.1996; \  E.L.Bolda, R.Y.Chiao,
J.C.Garrison, ``Two theorems for the group velocity in dispersive
media", {\em Phys. Rev., A}, vol.48, pp.3890-3894, Nov.1993; \
B.Macke, J.L.Queva, F.Rohart and B.Segard, ``Linear pulse propagation in
a resonant medium: The adiabatic limit", {\em J. de Physique}, vol.48,
pp.797-808, May 1987; \ C.G.B.Garret and D.E.McCumber, ``Propagation of a
Gaussian light pulse through an anomalous dispersion medium", {\em Phys. Rev.
A}, vol.1, pp.305-313, Feb.1970. \ See also C.R.Leavens and R.Sala-Mayato,
``Are predicted superluminal tunneling times an artifact of using the
nonrelativistic Schr\"odinger equation?", {\em Ann. der Physik}, vol.7(7-8),
pp.662-670, 1998; \ M.G.Muga and J.P.Palao, ``Negative times delays in
onedimensional absorptive collisions", {\em Ann. der Physik}, vol.7(7-8),
pp.671-678, 1998; \ J.G.Muga, I.L.Egusquiza, J.A.Damborenea and F.Delgado,
``Bounds and enhancenments for negative scattering time delays",
{\em Phys. Rev., A}, vol.66, 042115, Oct.2002 (8 pages); \ A.M.Shaarawi,
B.H.Tawfik and I.M.Besieris, ``Superluminal advanced transmission of X-waves
undergoing frustrated total internal reflection: The evanescent fields and the
Goos-H\"anchen effect", {\em Phys. Rev., E}, vol.66, 046626, oct.2002 (11
pages); \ M.A.Porras, I.Gonzalo and A.Mondello, ``Pulsed light beams in
vacuum with superluminal and negative group velocities", e-print physics/0202069,
Feb.2002; \ V.Petrillo and L.Refaldi: "A time asymptotic expression for the
wave function emerging from a quanto-mechanical barrier", subm. for pub., 2002; \
V.Petrillo, (in prepatation).\hfill\break

[30] L.J.Wang, A.Kuzmich and A.Dogariu, ref.[23]; \
M.W.Mitchell and R.Y.Chiao, ``Negative
group delay and ``fronts" in a causal system: An experiment with
very low frequency bandpass amplifiers", {\em Phys. Lett., A},
vol.230, pp.133-138, Jun.1997; \ B.Segard and B.Macke, ``Observation of
negative velocity pulse propagation", {\em Phys. Lett., A}, vol.109,
pp.213-216, May 1985; \ S.Chu and W.Wong, ``Linear
Pulse Propagation in an Absorbing Medium", {\em Phys. Rev.
Lett.}, vol.48, 738-741, March 1982. \ See also
G.Kurizki, A.Kozhekin, A.G.Kofman and M.Blaauboer, ref.[22]; \
G.Nimtz, ref.[22].\hfill\break

[31] G.Alzetta, A.Gozzini, L.Moi and G.Orriols, ``An
experimental method for the observation of the RF transitions and
laser beat resonances in oriented  Na  vapor", {\em Nuovo Cimento
B}, vol.36, pp.5-20, 1976.\hfill\break

[32] M.Artoni, G.C.La Rocca, F.S.Cataliotti and F.Bassani,
``Highly anomalous group velocity of light in ultracold Rubidium
gases", {\em Phys. Rev., A}, vol.63, 023805, 2001 (6 pages).\hfill\break

[33] H.Bateman, {\em Electrical and Optical Wave Motion} (Cambridge
Univ.Press; Cambridge, 1915); \ R.Courant and D.Hilbert, {\em Methods of
Mathematical Physics} (J.Wiley; New York, 1966), vol.2, p.760; \
J.N.Brittingham, ``Focus waves modes in
homogeneous Maxwell's equations: Transverse electric mode", {\em
J. Appl. Phys.}, vol.54, pp.1179-1189, March 1983; \
R.W.Ziolkowski, ``Exact solutions of the wave equation with complex
source locations", {\em J. Math. Phys.}, vol.26, pp.861-863,
Apr.1985; \ J.Durnin, J.J.Miceli and J.H.Eberly,
``Diffraction-free beams", {\em Phys. Rev. Lett.}, vol.58,
pp.1499-1501, Apr.1987; \ A.O.Barut, ``$E=\hbar\om$", {\em Phys.
Lett., A}, vol.143, pp.349-352, Jan.1990; \ A.O.Barut and
A.Grant, ``Quantum particle-like configurations of the
electromagnetic field", {\em Found. Phys. Lett.}, vol.3,
pp.303-310, Aug.1990; \ A.O.Barut and A.J.Bracken,
``Particle-like configurations of the electromagnetic field: an
extension of de Broglie's ideas", {\em Found. Phys.}, vol.22,
pp.1267-1285, Oct.1992; \ P.Hillion, ``Generalized Phases and
Nondispersive waves", {\em Acta Applicandae Matematicae},
vol.30, pp.35-45, Jan.1993.\hfill\break

[34] J.A.Stratton, {\em Electromagnetic Theory} (McGraw-Hill; New York,
1941), p.356; \ A.O.Barut and H.C.Chandola, ``Localized
tachyonic wavelet solutions of the wave equation", {\em Phys.
Lett., A}, vol.180, pp.5-8, Aug.1993; \ A.O.Barut, ``Localized
rotating wavelets with half integer spin", {\em Phys. Lett., A},
vol.189, pp.277-280, June 1994.\hfill\break

[35] R.Donnelly and R.W.Ziolkowski, ``Designing localized waves",
{\em Proc. Roy. Soc. London A}, vol.440, pp.541-565, March 1993; \
I.M.Besieris, A.M.Shaarawi and R.W.Ziolkowski, ``A bi-directional
traveling plane wave representation of exact solutions of the
scalar wave equation", {\em J. Math. Phys.}, vol.30, pp.1254-1269,
June 1989; \ S. Esposito, ``Classical solutions of Maxwell's equations
with group-velocity different from $c$, and the photon tunneling
effect", {\em Phys. Lett., A}, vol.225, pp.203-209, Feb.1997; \
J.Vaz and W.A.Rodrigues, {\em Adv. Appl. Cliff. Alg.},
vol.S-7, 457 (1997); \ V.V.Borisov and A.P.Kiselev, ``A new class of
relatively undistorted progressive ways", {\em Appl. Math. lett.},
vol.13, pp.83-86, 2000.\hfill\break

[36] See also E.Recami and W.A.Rodrigues Jr., ``A model theory
for tachyons in two dimensions", in {\em Gravitational Radiation
and Relativity}, J.Weber and T.M.Karade editors (World Scient.;
Singapore, 1985), pp.151-203, and refs. therein.\hfill\break

[37] A.M.Shaarawi, I.M.Besieris and R.W.Ziolkowski, ``A
novel approach to the synthesis of nondispersive wave packet
solutions to the Klein-Gordon and Dirac equations", {\em J. Math.
Phys.}, vol.31, pp.2511-2519, Oct.1990, Sect.VI; \ ``A space-time
representation of a massive, relativistic, spin-zero particle,
{\em Nucl. Phys. (Proc.Suppl.) B}, vol.6, 255-258, 1989; \
``Diffraction of a nondispersive wave-packet in the 2 slit
interference experiment", {\em Phys. Lett., A}, vol.188, pp.218-224,
May 1994. \ See also V.K.Ignatovich, ``Non-spreading wave packets
in quantum-mechanics", {\em Found. Phys.}, vol.8, pp.565-571, 1978;
and A.O.Barut, ``Formulation of wave mechanics without the Planck
constant h," {\em Phys. Lett., A}, vol.171, pp.1-2, Nov.1992; \ {\em Ann.
Foundation L. de Broglie}, Jan.1994; \ and ``Quantum
theory of single events, Localized de Broglie--wavelets, Schroedinger waves
and classical trajectories", preprint IC/90/99 (ICTP; Trieste, 1990).\hfill\break

[38] A.O.Barut, G.D.Maccarrone and E.Recami, ``On the shape
of tachyons", {\em Nuovo Cimento A}, vol.71, pp.509-533,
Oct.1982; \ P.Caldirola, G.D.Maccarrone and E.Recami, ``Second
contribution on solving the imaginary quantities problem in
Superluminal Lorentz transformations", {\em Lett. Nuovo Cimento},
vol.29, pp.241-250, Oct.1980; \ E.Recami and G.D.Maccarrone,
``Solving the imaginary quantities problem in Superluminal
Lorentz transformations", {\em Lett. Nuovo Cimento}, vol.28,
pp.151-157, May 1980. \ See also E.Recami refs.[3,4,42].\hfill\break

[39] J.-Y.Lu and J.F.Greenleaf, ``Nondiffracting X-waves: Exact
solutions to free-space scalar wave equation and their finite
aperture realizations", {\em IEEE Trans. Ultrason. Ferroelectr.
Freq. Control}, vol.39, pp.19-31, Jan.1992.\hfill\break

[40] J.Durnin, J.J.Miceli and J.H.Eberly, ref.[33]; \
J.Durnin, J.J.Miceli and J.H.Eberly, ``Comparison of
Bessel and Gaussian beams", {\em Opt. Lett.}, vol.13, pp.79-80,
Jan.1988.\hfill\break

[41] J.-Y.Lu and J.F.Greenleaf, ``Experimental verification of
nondiffracting X-waves", {\em IEEE Trans. Ultrason. Ferroelectr.
Freq. Control}, vol.39, pp.441-446, May 1992: In this case the
beam speed is larger than the {\em sound} speed in the considered
medium.\hfill\break

[42] E.Recami, ``On localized X-shaped Superluminal solutions to
Maxwell equations", {\em Physica A}, vol.252, pp.586-610,
Apr.1998; \ J.-Y.Lu, J.F.Greenleaf and E.Recami,
``Limited-Diffraction Solutions to Maxwell (and Schroedinger)
Equations", e-print physics/9610012 (Oct.1996). \ See also R.W.Ziolkowski,
I.M.Besieris and A.M.Shaarawi, ``Aperture realizations of
exact solutions to homogeneous wave-equations", {\em J. Opt. Soc.
Am., A}, vol.10, pp.75-87, Jan.1993.\hfill\break

[43] P.Saari and K.Reivelt, ``Evidence of X-shaped
propagation-invariant localized light waves", {\em Phys. Rev.
Lett.}, vol.79, pp.4135-4138, Nov.1997.\hfill\break

[44] D.Mugnai, A.Ranfagni and R.Ruggeri, ``Observation of
superluminal behaviors in wave propagation", {\em Phys. Rev.
Lett.}, vol.84, pp.4830-4833, May 2000: this paper
aroused some criticisms, to which the authors replied. \
P.Di Trapani, G.Valiulis, A.Piskarskas, O.Jedrkiewicz, J.Trull,
C.Conti and S.Trillo, ``Spontaneous formation of nonspreading
X-shaped wavepackets", Dec.2000 (submitted to {\em Science}).\hfill\break

[45] M.Z.Rached, E.Recami and H.E.Hern\'andez F., ``New
localized Superluminal solutions to the wave equations with
finite total energies and arbitrary frequencies" [e-print physics/0109062],
{\em Eur. Phys. J., D}, vol.21, pp.217-228, Sept.2002; \
M.Z.Rached, E.Recami and F.Fontana, ``Superluminal localized solutions to
Maxwell equations propagating along a waveguide: The finite-energy case"
[e-print physics/0209102], submitted to {\em Phys. Rev., E}. \ Cf.
also, e.g., M.A.Porras,
R.Borghi and M.Santarsiero, ``Superluminality in Gaussian beams",
{\em Opt. Commun.}, vol.203, pp.183-189, March 2002.\hfill\break

[46] M.Z.Rached, E.Recami and F.Fontana, ``Superluminal
localized solutions to Maxwell equations propagating along a
normal-sized waveguide" [e-print physics/0001039],
{\em Phys. Rev., E}, vol.64, 066603, Dec.2001 (6 pages); \  M.Z.Rached,
K.Z.N\'obrega, E.Recami and H.E.Hernandez F., ``Superluminal
X-shaped beams propagating without distortion along a coaxial
guide", {\em Phys. Rev., E}, vol.66, 046617, Oct.2002 (ten pages); \
I.M.Besieris, M.Abdel-Rahman, A.Shaarawi and A.Chatzipetros, ``Two
fundamental representations of localized pulse solutions of the scalar
wave equation", {\em Progress in Electromagnetic Research (PIERS)}, vol.19,
pp.1-48, 1998.\hfill\break

[47] J.-y.Lu, H.-h.Zou and J.F.Greenleaf, ``Biomedical
ultrasound beam forming", {\em Ultrasound in Medicine and
Biology}, vol.20, pp.403-428, 1994; \ J.-y.Lu and
J.F.Greenleaf, ``Producing deep depth of field and depth-independent
resolution in NDE with limited diffraction beams", {\em
Ultrasonic Imaging}, vol.15, pp.134-149, Apr.1993.\hfill\break

[48] M.Z.Rached and H.E.Hernandez-Figueroa, ``A rigorous
analysis of localized wave propagation in optical fibers", {\em
Opt. Comm.}, vol.191, pp.49-54, May 2000. \ From the
experimental point of view, cf. S.Longhi, P.Laporta, M.Belmonte
and E.Recami, ref.[25].\hfill\break

[49] P.Saari and H.S\~{o}najalg, ``Pulsed Bessel beams", {\em Laser
Phys.}, vol.7, pp.32-39, Jan.1997.\hfill\break

[50] Similar solutions were considered in A.T.Friberg, J.Fagerholm
and M.M.Salomaa, ``Space-frequency analysis of
nondiffracting pulses", {\em Opt. Commun.}, vol.136, pp.207-212,
March 1997; and J.Fagerholm, A.T.Friberg, J.Huttunen,
D.P.Morgan and M.M.Salomaa, ``Angular-spectrum representation of
nondiffracting X waves", {\em Phys. Rev., E}, vol.54, pp.4347-4352,
Oct.1996; as well as in P.Saari, in {\it Time's Arrows, Quantum Measurements
and Superluminal Behavior}, D.Mugnai et al. editors (C.N.R.; Rome, 2001),
pp.37-48.\hfill\break

[51] I.S.Gradshteyn and I.M.Ryzhik, {\em Integrals, Series and Products},
4th edition (Ac.Press; New York, 1965).\hfill\break

[52] In the luminal case ($V=c$), such pulses have been studied in
R.W.Ziolkowski, ``Localized transmission of electromagnetic
energy", {\em Phys. Rev., A}, vol.39, pp.2005-2033, Feb.1989; \
A.Shaarawi, I.M.Besieris and R.W.Ziolkowski, ref.[35]; \
I.M.Besieris, M.Abdel-Rahman, A.Shaarawi and A.Chatzipetros,
ref.[46].\hfill\break

[53] I.M.Besieris, M.Abdel-Rahman, A.Shaarawi and A.Chatzipetros,
ref.[46].\hfill\break

[54] R.M.Herman and T.A.Wiggins, ``Production and uses of
diffractionless beams", {\em J. Opt. Soc. Am., A}, vol.8,
pp.932-942, June 1991; \ H.S\~{o}najalg, A.Gorokhovskii, R.Kaarli,
V.Palm, M.Ratsep, P.Saari, ``Optical pulse shaping by filters
based on spectral hole burning", {\em Opt. Commun.}, vol.71,
pp.377-380, Jun.1989.\hfill\break

[55] G.P.Agrawal: {\em Nonlinear Fiber Optics}, 2nd edition (Ac.Press;
New York, 1995). \hfill\break

[56] Cf., e.g., also M.Zamboni-Rached, K.Z.N\'obrega, E.Recami and
H.E.Hern\'andez F., ref.[46].

\end{document}